\documentclass[sigconf,pbalance]{arxivstyle} %

\usepackage{xcolor}  %

\microtypecontext{spacing=nonfrench}

\usepackage[english]{babel}

\usepackage[most]{tcolorbox} %

\usepackage{mathtools}

\usepackage{algpseudocode}
\usepackage{amsmath}
\usepackage{amssymb}
\usepackage{amsthm}
\usepackage{algorithm}

\usepackage{graphicx}
\usepackage{subfigure}
\usepackage{tikz}
\usetikzlibrary{positioning, arrows, arrows.meta, shapes.geometric, fit, backgrounds}

\DeclareRobustCommand{\circled}[1]{\tikz[baseline=(char.base)]{
    \node[shape=circle,draw,inner sep=1pt,font=\footnotesize] (char) {#1};}}

\usepackage{colortbl}
\usepackage{array}
\usepackage{tabularx}
\usepackage{pifont} %
\usepackage{xcolor} %
\usepackage{booktabs}
\usepackage{varwidth} %
\usepackage{multirow}

\usepackage{rotating}   %

\usepackage{xspace}
\usepackage{siunitx}
\definecolor{checkgreen}{RGB}{50,200,50}
\definecolor{crossred}{RGB}{255,50,50}
\newcommand{\cmark}{\textcolor{checkgreen}{\ding{51}}}
\newcommand{\xmark}{\textcolor{crossred}{\ding{55}}}

\usepackage{cleveref}
\crefname{section}{Section}{Sections}
\crefname{figure}{Figure}{Figures}
\crefname{table}{Table}{Tables}
\crefname{equation}{Equation}{equations}
\crefname{algocf}{Algorithm}{algorithms}
\crefname{appendix}{Appendix}{appendices}

\usepackage{tcolorbox}

\usepackage{bbding}

\usepackage{enumerate}
\usepackage{enumitem}

\usepackage[misc]{ifsym}

\usepackage{wasysym}
\setcopyright{cc} %
\setcctype{by}
\copyrightyear{2026} %
\acmYear{2026} %
\acmDOI{10.1145/3830454.3846592} %
\acmConference[CCS '26]{Proceedings of the 2026 ACM SIGSAC Conference on Computer and Communications Security}{November 15--19, 2026}{The Hague, Netherlands}
\acmBooktitle{Proceedings of the 2026 ACM SIGSAC Conference on Computer and Communications Security (CCS '26), November 15--19, 2026, The Hague, Netherlands}
\acmISBN{979-8-4007-2871-6/2026/11} %
\newcommand{\itemtext}[1]{#1}
\begin{document}
\title[Mind the Gap: Cross-RFC Formal Analysis of DNSSEC]{Mind the Gap: Cross-RFC Formal Analysis Exposes \\
DNSSEC's Hidden Vulnerabilities}

\author{Qifan Zhang}
\orcid{0000-0002-9278-9576}
\authornote{Most of Qifan Zhang's work was done when he was a student in UC Irvine.}
\affiliation{%
  \institution{Palo Alto Networks}
  \city{Santa Clara}
  \state{California}
  \country{USA}}
\email{qzhang@paloaltonetworks.com}

\author{Zilin Shen}
\orcid{0000-0001-5342-0998}
\affiliation{%
  \institution{Purdue University}
  \city{West Lafayette}
  \state{Indiana}
  \country{USA}}
\email{shen624@purdue.edu}

\author{Imtiaz Karim}
\orcid{0009-0000-8680-9932}
\affiliation{%
  \institution{The University of Texas at Dallas}
  \city{Richardson} 
  \state{Texas}
  \country{USA}}
\email{imtiaz.karim@utdallas.edu}

\author{Elisa Bertino}
\orcid{0000-0002-4029-7051}
\affiliation{%
  \institution{Purdue University}
  \city{West Lafayette}
  \state{Indiana}
  \country{USA}}
\email{bertino@purdue.edu}

\author{Daiping Liu}
\orcid{0000-0002-9660-4444}
\affiliation{%
  \institution{Palo Alto Networks}
  \city{Santa Clara}
  \state{California}
  \country{USA}}
\email{dpliu@paloaltonetworks.com}

\author{Zhou Li}
\orcid{0000-0002-9401-1012}
\authornote{Corresponding author.}
\affiliation{%
  \institution{University of California, Irvine}
  \city{Irvine}
  \state{California}
  \country{USA}}
\email{zhou.li@uci.edu}

\renewcommand{\shortauthors}{Qifan Zhang et al.}

\begin{abstract}
This paper presents a cross-RFC formal analysis of selected DNSSEC mechanisms using SAPIC+ models of cryptographic operations and stateful caching. We analyze properties concerning data origin authenticity, integrity, chain of trust, and authenticated denial of existence within these models. Our analysis shows that core authentication properties hold under the modeled assumptions, while \emph{wildcard expansion}, the mechanism that bridges positive and negative authentication, exhibits a cross-RFC interaction vulnerability that defeats DNSSEC's integrity guarantee.
Specifically, we discover \emph{Cross-Zone Wildcard Cache Poisoning}: a new attack in which a malicious zone operator exploits a gap in the DNSSEC specifications to poison resolver caches with forged records for some domains outside its authority, served to clients as DNSSEC-authentic. The tested versions of all four major resolvers implementing the affected caching optimization accept the malformed wildcard material. The resulting impact depends on the resolver's synthesis path. All four vendors have publicly released fixes. The distinctive integrity failure is a synthesized, attacker-chosen positive answer with the Authenticated Data (AD) bit set: prior attacks instead exhausted availability, downgraded to AD$=$0, or skipped validation.
Our models also identify two cross-RFC availability flaws and rediscover three known attack classes. We test six resolvers and measure deployment prerequisites through focused resolver and TLD scans. We propose updates to four RFCs.
\end{abstract}



\maketitle

\section{Introduction}
\label{sec:intro}

The Domain Name System (DNS)~\cite{rfc2181} translates domain names into IP addresses.
DNS Security Extensions (DNSSEC)~\cite{rfc4033,rfc4034,rfc4035} authenticate DNS responses using cryptographic signatures.
However, researchers have discovered vulnerabilities in DNSSEC that can cause man-in-the-middle attacks~\cite{heftrig2023downgrading} and denial-of-service attacks~\cite{heftrig2024harder, gruza2024attacking} by analyzing DNS implementations.
While identifying and mitigating these vulnerabilities strengthens the DNSSEC ecosystem, so far there has been no formal analysis of the DNSSEC specifications that models the cross-RFC interactions governing resolver behavior to assess the security of DNSSEC \textit{from its core}. 
Such a research gap casts doubt on the security guarantees provided by DNSSEC~\cite{chung2017longitudinal} and impedes its broader adoption by DNS resolvers, users, and domain owners~\cite{chung2017understanding, le2018economic}.

\vspace{2pt} \noindent \textbf{Prior Works and Challenges.}
Previous research on DNSSEC analysis has looked into scattered aspects.
Empirical studies including Downgrading DNSSEC and KeyTrap~\cite{heftrig2023downgrading,heftrig2024harder} analyze RFC language as well as resolver behavior, tracing vulnerabilities to specification ambiguities and design choices. Ariyapperuma and Mitchell~\cite{ariyapperuma2007security} also analyze DNSSEC design weaknesses. Our contribution is an executable symbolic analysis of selected cross-RFC validation and caching interactions. Finite-state and theorem-proving efforts were applied to analyze DNSSEC formally~\cite{bau2010security,kammuller2014verification,eixarch2012formal} but they omitted cryptographic reasoning. Chetioui et al.~\cite{chetioui2019formal} verified only a single property with ProVerif, while DNS-Maude~\cite{liu2023formal} and Zheng et al.~\cite{zheng2023automated} formalized plain DNS resolution without modeling DNSSEC validation. A detailed technical comparison appears in Section~\ref{sec:related}.

Formal verification of DNSSEC faces three challenges. First, DNSSEC is not defined by a single document: a resolver processing one query must simultaneously obey many RFCs (core validation, denial of existence, caching, operational extensions) that were written independently yet govern overlapping steps of the same resolution path. We use \emph{cross-RFC interaction} to mean behavior mandated by one RFC interacting with assumptions made by another on that path. Under-specified states and transitions at RFC boundaries complicate this analysis.
Second, the resolver cache is \textit{stateful}: an entry's trustworthiness depends on the validation context at insertion time, which may later diverge from the resolver's current state. Prior formal models such as DNS-Maude~\cite{liu2023formal} do not track these security-relevant state changes under DNSSEC.
Third, DNSSEC admits alternative cryptographic mechanisms for the same request type whose proofs operate over incompatible \textit{name-ordering domains}: NSEC reasons about lexicographic intervals over owner names, whereas NSEC3 reasons about intervals over hashed names, so the same queried name falls into different ``coverage'' ranges under each mechanism. Both orderings require custom coverage predicates in our symbolic model. %

Denial-of-existence, or \emph{negative authentication}, compounds these challenges. Positive authentication signs existing records, whereas NSEC and NSEC3 prove absence using ordered names or hashes.
Reusing proofs under NTA (RFC~7646~\cite{rfc7646}) and aggressive caching (RFC~8198~\cite{rfc8198}) requires tracking validation context.

\vspace{2pt} \noindent \textbf{Our Goal and Design.}
In this work, we present a cross-RFC formal analysis of DNSSEC specifications. We systematically analyzed \textit{10} RFCs related to DNSSEC and identified \textit{20} security-related properties covering core DNSSEC protocols including DNS message exchange, static ZSK/KSK relationships, and chain-of-trust validation, as well as auxiliary modes like Negative Trust Anchors (NTA) and aggressive caching. We additionally formalize \textit{3} cross-RFC interaction properties that capture cross-RFC interactions explicitly (23 properties total, see Table~\ref{tab:verification-results}).
To correctly verify cache states under DNSSEC, we develop a new two-dimensional cache state tracking model, covering both freshness and validation context, enabling the detection of cache entries that outlive their security guarantees. Finally, to address the third challenge of incompatible name-ordering domains, we introduce custom symbolic predicates that approximate NSEC's lexicographic ordering, NSEC3's hash-based ordering, and wildcard coverage, allowing denial-of-existence reasoning to be plugged directly into off-the-shelf symbolic verifiers. %

Since DNSSEC relies on cryptographic functions, 
our models are written in SAPIC+~\cite{cheval2022sapic+} and translated to Tamarin~\cite{meier2013tamarin} and ProVerif~\cite{blanchet2018proverif}, two symbolic verifiers.
SAPIC+ supports the stateful and cryptographic models needed for these properties~\cite{cheval2022sapic+,jacomme2023comprehensive}.

\vspace{2pt} \noindent \textbf{Results.}
Within the scope of our models, we establish properties concerning data origin authenticity, data integrity, chain of trust, authenticated denial of existence, and cache correctness. Six of the 23 analyzed properties are violated (Table~\ref{tab:verification-results}).
Our models also rediscovered three known attack classes (zone enumeration~\cite{wala2024zone}, algorithm downgrade~\cite{heftrig2023downgrading}, CD-bit cache poisoning~\cite{zhang2025your}), checking consistency with those known attacks.
We find that properties related to positive authentication (authenticity, integrity, chain of trust) hold when RFCs are followed individually,
while \emph{wildcard expansion}, the mechanism that bridges positive and negative authentication, exhibits a cross-RFC interaction vulnerability that defeats DNSSEC's integrity guarantee under aggressive caching.

Our survey identified wildcard processing and aggressive caching as a model-ready interaction (Section~\ref{subsec:rfc_survey}). The minimal model isolates the signer-bound invariant using selected names and abstracted RFC checks. A ProVerif counterexample exposes its violation. Adding a signer-depth guard excludes that construction (Section~\ref{subsec:model_arch}). Specifically, our analysis discovers \emph{Cross-Zone Wildcard Cache Poisoning} (Property~21 violation): RFC~4034 \S3.1.3 upper-bounds the RRSIG \texttt{Labels} field but RFC~4035 \S5.3.2 reconstructs the wildcard owner from \texttt{Labels} without lower-bounding it against the signer's zone depth. A malicious child zone can therefore craft a cryptographically valid RRSIG whose reconstructed wildcard owner lies in the \emph{parent} zone. Under RFC~8198 wildcard synthesis, the resolver then serves forged A records for some sibling domains with the AD (Authenticated Data) bit set. The missing bound appears in the tested BIND, Unbound, Knot Resolver, and PowerDNS versions, but serving outcomes differ: Unbound revalidates and blocks the basic NSEC trace, whereas the demonstrated AD=1 outcomes use vulnerable synthesis paths (Section~\ref{subsec:pitfalls}). All four vendors have publicly released fixes. Unlike prior DNSSEC attacks, this requires no on-path position, no operator misconfiguration, and no key compromise: any registrant under an NSEC-signed parent that publishes an NSEC covering the target name can poison some sibling domains at a vulnerable downstream resolver with RFC~8198 wildcard synthesis enabled. The attack requires a covering parent NSEC and does not overwrite arbitrary existing parent-zone RRsets.
To our knowledge, this is the first protocol-level attack that yields attacker-chosen positive answers with AD$=$1 through DNSSEC wildcard synthesis. KeyTrap~\cite{heftrig2024harder} causes CPU exhaustion during validation. Downgrading DNSSEC~\cite{heftrig2023downgrading} produces forged data with AD$=$0 after validation is skipped. Here, the missing signer-depth bound allows a valid child-zone signature to authenticate wildcard data that a vulnerable resolver later uses outside that zone.
Section~\ref{subsec:additional_findings} describes two further cross-RFC availability flaws: NTA cache poisoning and NSEC/NSEC3 coexistence. Their conditions and resolver-dependent outcomes appear in Table~\ref{tab:aggressive_vuln}, with further analysis in Appendix~\ref{app:additional_vulns}.
Separate measurements characterize DNSSEC deployment and the prerequisites for P21 (Section~\ref{subsec:pitfalls}). We propose four RFC updates to address these gaps (Section~\ref{sec:rootcause}).

\vspace{2pt} \noindent \textbf{Contributions.} Our contributions are summarized below:

\begin{itemize}[noitemsep,nolistsep]
\item We present a cross-RFC formal analysis of DNSSEC, modeling cross-RFC interactions under various setups (e.g., wildcard expansion, NTA, and aggressive caching).
We formally check 20 baseline and 3 cross-RFC interaction security properties (23 total). The 3 cross-RFC interaction properties are situated within a systematic survey of 52 cross-RFC interaction points in a selected 41-RFC DNSSEC corpus, with one extension added during review (Section~\ref{subsec:rfc_survey}).

\item We discover three novel vulnerability classes arising from cross-RFC gaps, led by \emph{Cross-Zone Wildcard Cache Poisoning} (Section~\ref{subsec:pitfalls}).
Two additional DoS-class issues (NTA Cache Poisoning and NSEC/NSEC3 Coexistence) were also identified.
We validate all three through targeted testing of six major DNS resolver implementations.

\item \itemtext{We propose updates to RFC~4035, RFC~7646, RFC~8198, and RFC~5155. A focused resolver scan and complete TLD census measure separate P21 prerequisites.}

\end{itemize}

This extended version adds model details, historical measurement tables, RFC-interaction listings, and additional attack analysis to the CCS~2026 paper (Appendices~\ref{sec:append_model}--\ref{app:additional_vulns}). It will be released on arXiv\footnote{\url{https://doi.org/10.48550/arXiv.2512.11431}}.
\section{Background}
\label{sec:background} 

\subsection{DNS and DNSSEC}
\label{subsec:dns_dnssec}

The Domain Name System (DNS) is a hierarchical naming infrastructure that translates domain names into IP addresses.
DNS queries traverse authoritative nameservers (NSes) organized hierarchically: root servers, Top-Level Domain (TLD) servers, and domain-specific authoritative servers. Each NS hosts a zone file containing resource records (\textbf{\texttt{RR}}). The original DNS lacked security guarantees, leaving it vulnerable to cache poisoning~\cite{kaminsky2008black}.

\begin{figure}[ht!]
\centering
\includegraphics[page=1,width=1\columnwidth]{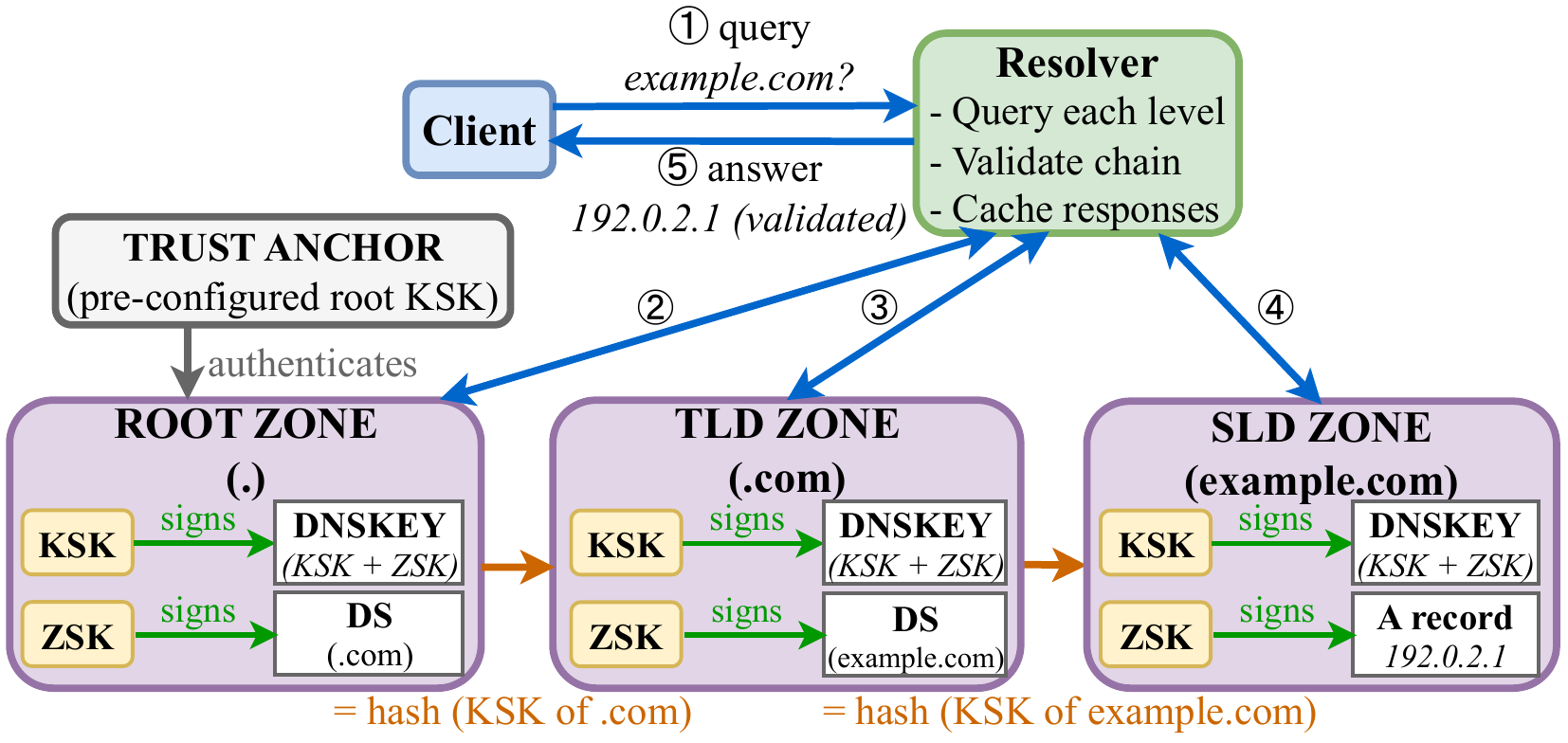}

\caption{
The DNSSEC chain of trust and resolution process: When a client queries a domain, the resolver iteratively queries authoritative nameservers at each hierarchy level, validating responses using the chain of trust. }
\label{fig:DNSSEC}
\Description{Hierarchy and message-flow diagram connecting a client and resolver to three authoritative zones. Step 1 sends a query for example.com to the resolver. Steps 2, 3, and 4 connect the resolver to the root, com, and example.com zones, respectively. A preconfigured root key-signing key provides the trust anchor. Within each zone, the key-signing key signs the DNSKEY set containing the key-signing and zone-signing keys. The root and com zone-signing keys sign DS records linking to the next zone's key-signing key. The example.com zone-signing key signs an A record containing 192.0.2.1. Step 5 returns that validated address to the client. Numbered bidirectional arrows represent DNS exchanges. Separate signing and delegation arrows show the cryptographic trust links.}
\end{figure}

DNS Security Extensions (DNSSEC) address these security gaps by using public-key cryptography to sign DNS resource records (RRs), allowing resolvers to verify the source and correctness of DNS responses (Figure~\ref{fig:DNSSEC}). The main records under DNSSEC include: 

\vspace{2pt} \noindent \textbf{\texttt{DNSKEY} Records} store public keys used to sign and validate RRs within a zone that is associated with a domain name. Each zone typically has a Key Signing Key (KSK) and a Zone Signing Key (ZSK). These keys must be periodically replaced (\textit{rolled over}) to limit exposure from potential key compromise, a process requiring careful coordination to maintain an unbroken chain of trust.

\vspace{2pt} \noindent \textbf{\texttt{RRSIG} Records} contain the cryptographic signature of a set of RRs that share the same name, type, and class (called \textbf{\texttt{RRSet}}). Resolvers verify the signature using the ZSK from the corresponding \texttt{DNSKEY} record. The RRSIG also carries a \texttt{Labels} field (RFC~4034 \S3.1.3) recording the label count of the original owner name, which validators use to reconstruct the canonical owner when the answer was synthesized from a wildcard.

\vspace{2pt} \noindent \textbf{\texttt{DS} Records} serve as a \textit{``pointer''} record stored in the parent zone that references the child zone's KSK. This establishes a \textit{chain of trust} from the parent zone to the child zone. 

\vspace{2pt} \noindent \textbf{\texttt{NSEC}/\texttt{NSEC3} Records} prove the non-existence of a requested name or record type through cryptographic proofs linking adjacent domain names. While both mechanisms prove non-existence, they employ different ordering systems with security implications we analyze in Section~\ref{subsec:pitfalls}.

DNS record validation under DNSSEC (or DNSSEC PKI) relies on creating a chain of trust starting from the root zone (signed by a root KSK) and propagating signatures down to each child zone through the use of DS records. This design allows resolvers to validate a DNS response by tracing signatures upward in the domain hierarchy until reaching a known trust anchor (the root KSK or any other configured trust anchor). By validating each signature in the chain, DNSSEC resolvers can detect alterations or forgeries.

\noindent \textbf{DNSSEC-related Auxiliary Modes.}  
To reduce the failure rate of DNSSEC validation in real-world deployment, a few auxiliary modes were proposed, which can be turned on by resolver administrators. We focus on two auxiliary modes defined in two RFCs. 

First, RFC~7646~\cite{rfc7646} defines Negative Trust Anchors (NTA) to temporarily disable DNSSEC validation for a specific misconfigured domain name. It also comes with a mandatory expiration (``NTAs MUST expire automatically when their configured lifetime ends'') which ``SHOULD NOT exceed a week''. Second, RFC~8198~\cite{rfc8198} specifies aggressive caching for NSEC/NSEC3 records to allow DNSSEC-validating resolvers to generate negative answers within a range and positive answers from wildcards, decreasing latency and increasing privacy. 
These two auxiliary modes are the model-ready edges identified by the cross-RFC survey of Section~\ref{subsec:rfc_survey}, and their interactions with core DNSSEC validation yield the vulnerabilities described in Section~\ref{subsec:pitfalls}.

\subsection{SAPIC+}
\label{subsec:tamarin}

DNSSEC protocols rely on cryptography to prove authenticity and integrity for DNS records.  
To analyze network protocols with cryptographic components, Tamarin~\cite{meier2013tamarin} and ProVerif~\cite{blanchet2018proverif} are widely used tools for formal analysis. 
SAPIC+~\cite{cheval2022sapic+} was developed as a unified verification platform that integrates multiple backend verifiers, including Tamarin~\cite{meier2013tamarin} and ProVerif~\cite{blanchet2018proverif}. This lets us draw on the strengths of each tool: Tamarin for stateful protocols with complex temporal properties, and ProVerif for faster automated proofs of simpler properties. SAPIC+'s correctness proof also enables the reuse of results proved in one tool across other tools~\cite{jacomme2023comprehensive}. Here we describe how protocols are specified in SAPIC+.

\noindent \textbf{SAPIC+ Syntax.}  
As common in symbolic models, messages are represented as \emph{terms}, which can be used to model fresh values, messages, and variables.  
Cryptographic operations are represented using \emph{function symbols}. For example, $\mathsf{sign}(m,k)$ is a signing function with arity 2.  
Equations define the primitives represented by function symbols. For example,
\[
\mathsf{verify}(\mathsf{sign}(m,k),m,\mathsf{pk}(k)) = \mathsf{true}
\] 
\noindent states that the verification of a message $m$ with the private signing key $k$ is successful if the signature matches the message and the corresponding public key $\mathsf{pk}(k)$.  
This equation is applied during the signature verification of DNSSEC.

\noindent \textbf{Roles in SAPIC+.}  
Protocols are modeled as processes in a dialect of the applied pi-calculus~\cite{abadi2017applied}, a language for modeling cryptographic protocols as parallel and communicating programs.  
DNSSEC roles are SAPIC+ processes communicating over a network controlled by a Dolev-Yao adversary~\cite{dolev1983security}, which can intercept and inject messages. A client process emits $\langle\texttt{qname},\texttt{qtype}\rangle$ on a channel shared with the resolver. Appendix~\ref{sec:append_model} summarizes the encoding. The artifact contains the process sources (Appendix~\ref{app:open_science}).
\section{Cross-RFC Analysis of DNSSEC}
\label{sec:formal_models}

\begin{table}[ht]
\small
\centering
\caption{DNSSEC-relevant RFCs represented in our SAPIC+ models. Section~\ref{subsec:rfc_survey} describes selection from the 41-RFC corpus. The artifact records modeled clauses and abstractions.
}
\label{tab:dnssec-rfcs}
\setlength{\tabcolsep}{4pt}
\begin{tabular}{ll}
\toprule
\textbf{RFC} & \textbf{Role} \\
\midrule
4033~\cite{rfc4033} & Core: introduction \& requirements \\
4034~\cite{rfc4034} & Core: DNSKEY/RRSIG/NSEC/DS RR types \\
4035~\cite{rfc4035} & Core: validator \& signer protocol \\
4592~\cite{rfc4592} & Wildcard semantics \\
5155~\cite{rfc5155} & NSEC3 hashed denial of existence \\
4641~\cite{rfc4641} & KSK/ZSK management (obs.\ by 6781) \\
6781~\cite{rfc6781} & DNSSEC operational practices v2 \\
7646~\cite{rfc7646} & Negative Trust Anchors \\
8198~\cite{rfc8198} & Aggressive use of DNSSEC-validated cache \\
8624~\cite{rfc8624} & Algorithm implementation requirements \\
\bottomrule
\end{tabular}
\end{table}

DNSSEC's security guarantees emerge not from individual RFCs, but from their \emph{cross-RFC interactions}. While each RFC specifies behavior within its scope, critical security assumptions often span RFC boundaries without explicit specification.
Section~\ref{subsec:rfc_survey} describes how we systematically survey the DNSSEC RFC corpus to enumerate these cross-RFC dependencies. Table~\ref{tab:dnssec-rfcs} lists the ten RFCs whose normative behavior we ultimately encode in the SAPIC+ model, covering core DNSSEC mechanisms (signing and verifying all record types described in Section~\ref{subsec:dns_dnssec}), negative authentication (NSEC/NSEC3), key management, chain of trust, and auxiliary modes such as NTA and aggressive caching. The artifact's \texttt{MODEL\_INDEX.md} maps model files to RFC sections and records abstractions and omitted behavior (Appendix~\ref{app:open_science}).

Our DNS message abstraction represents message sections and selected flags, including CD and AD, but omits lower-layer fields such as UDP headers. It follows the message-level abstraction used by DNS-Maude~\cite{liu2023formal}.

\vspace{2pt} \noindent \textbf{Threat Model.}
We adopt the Dolev-Yao adversary model~\cite{dolev1983security}, where attackers control the network and can eavesdrop, inject, modify, or replay DNS messages. The adversary \emph{can} intercept all network traffic between resolvers and authoritative nameservers, forge or modify DNS responses in transit, trigger arbitrary queries to resolvers, and serve records from adversary-controlled zones using the adversary's own valid cryptographic keys. The adversary \emph{cannot} compromise the private signing keys (ZSKs/KSKs) of honest zones, break cryptographic primitives, or directly access resolver internal state. For our headline finding, the delegated-zone attacker requires no on-path position and no operator misconfiguration: control of a signed child zone suffices when the parent publishes a covering NSEC and the downstream resolver has the vulnerable synthesis behavior. This model aligns with other works of protocol verification~\cite{basin2018formal, cremers2017comprehensive}.
In Section~\ref{subsec:pitfalls}, we detail the attacker positions and goals for each vulnerability.

\subsection{Surveying Cross-RFC Interactions}
\label{subsec:rfc_survey}

\newcolumntype{Q}[1]{>{\raggedright\arraybackslash}p{#1}}

\begin{table}[t]
\centering
\caption{Disposition of 52 cross-RFC edges: 51 from the 41-RFC corpus and the review-added RFC~3596 extension E5 (Section~\ref{subsec:rfc_survey}). The 14 model-ready edges follow the three patterns discussed in Section~\ref{subsubsec:composition_survey}.}
\label{tab:rfc_interactions}
\footnotesize
\setlength{\tabcolsep}{3pt}
\begin{tabular}{@{}Q{2.35cm}cQ{2.9cm}Q{2.0cm}@{}}
\toprule
\textbf{Class} & \textbf{No.} & \textbf{Example RFC pairs} & \textbf{Disposition} \\
\midrule
Covered (Properties~1--23) & 10 & 4035$\times$8198, 4592$\times$4035, 3225$\times$4035 & Table~\ref{tab:verification-results} \\
\addlinespace[1pt]
Model-ready (P21--P23 patterns) & 14 & 6840$\times$4035, 9276$\times$8198, 8767$\times$4035, 4470$\times$8198 & Section~\ref{subsubsec:composition_survey} \\
\addlinespace[1pt]
Extension (E1--E5) & 13 & 5011, 7344/8078, 9824, 6781/8901, 3596 & Section~\ref{sec:discussion} \\
\addlinespace[1pt]
Out of scope (signaling / registry / out-of-band) & 15 & 8145, 8509, 8914, 7477, 9718 & Section~\ref{subsec:rfc_survey} \\
\midrule
\textbf{Total} & \textbf{52} & \multicolumn{2}{l}{41-RFC corpus + E5} \\
\bottomrule
\end{tabular}
\end{table}

Our survey identifies clauses where one RFC constrains state or behavior defined by another. These interactions determine the model's scope and the recurring patterns examined in Section~\ref{subsec:aggressive_caching}.

\vspace{2pt} \noindent \textbf{Corpus.}
The survey covers DNSSEC-related RFCs from DNS Extensions (\texttt{dnsext})~\cite{ietf-dnsext-docs}, DNS Operations (\texttt{dnsop})~\cite{ietf-dnsop-docs}, and related publication streams. We retain an RFC if it (a)~defines or modifies a DNSSEC RR type (DS, DNSKEY, RRSIG, NSEC, NSEC3, NSEC3PARAM, CDS/CDNSKEY, CSYNC, ZONEMD), (b)~defines validator or signer behavior, or (c)~defines an operational mechanism that touches validation state (trust anchors, validation bypass, cache synthesis).
We exclude pre-DNSSEC-bis specifications, including RFC~2535, that RFC~4033--4035 obsolete. This filter yields a 41-RFC corpus.

\vspace{2pt} \noindent \textbf{Interaction Extraction.}
For each corpus RFC we extract every paragraph that contains both an RFC~2119 keyword (MUST, SHOULD, MAY, REQUIRED, RECOMMENDED) and a term describing validation or cache state (\emph{validator}, \emph{resolver}, \emph{cache}, \emph{trust anchor}, an RFC~4035 validation state, or a DNSSEC RR-type name), yielding 406 candidate paragraphs. We then manually tag each candidate with the external RFC whose state it constrains, producing what we call a \emph{cross-RFC edge}: a clause-level interaction between two RFCs. Distinct clauses of the same RFC pair can therefore contribute separate edges. The original survey identified 51 edges. We additionally list the reviewer-suggested RFC~5155/RFC~3596 operational-namespace interaction as extension E5, giving 52 entries in the revised inventory. E5 remains unmodeled.

\vspace{2pt} \noindent \textbf{Triage.}
Each edge is classified into one of four classes shown in Table~\ref{tab:rfc_interactions}.
\emph{Covered} (10 edges): the constraint is already expressed by one of our 23 properties (Table~\ref{tab:verification-results}). \emph{Model-ready} (14 edges): the constraint is expressible over the SAPIC+ model of Section~\ref{sec:design} without new state or actors. Section~\ref{subsubsec:composition_survey} shows that each instantiates one of three recurring failure patterns anchored by Properties~21--23. \emph{Extension}
(13 edges, groups E1--E5): requires extending the model with new state machines or actors, such as a trust-anchor lifecycle or parent-side CDS polling. These extensions are outside the modeled scope (Section~\ref{sec:discussion}). \emph{Out of scope} (15 edges): signaling, telemetry, registry-side, or out-of-band mechanisms that do not alter resolver validation or cache state. Every RFC listed in Table~\ref{tab:dnssec-rfcs} matches either source or sink of a \emph{covered} or \emph{model-ready} edge. Examples excluded from the model include trust-anchor telemetry, extended error reporting, and root trust-anchor distribution. Appendix~\ref{app:rfc_survey} retains supporting edge listings and exclusion details.
Table~\ref{tab:verification-results} summarizes the 23 security properties we verify (Section~\ref{sec:design}). The three cross-RFC interaction properties (P21--P23) instantiate the model-ready failure patterns above.
\newcolumntype{P}[1]{>{\raggedright\arraybackslash}p{#1}}

\begin{table*}[!htbp]
\centering
\caption{Formal Verification Results for DNSSEC with Cache Properties}
\label{tab:verification-results}
\footnotesize
\setlength{\tabcolsep}{4pt}
\begin{tabular*}{\textwidth}{@{\extracolsep{\fill}}clP{2.6cm}P{7.1cm}ccrl@{}}
\toprule
\textbf{\#} & \textbf{Analysis Group} & \textbf{Property Name} & \textbf{Property Description} & \textbf{Cache} & \textbf{Verified} & \textbf{Time} & \textbf{Tool} \\
\midrule

1 & \multirow{4}{*}{\shortstack[l]{Cache -- \\Core Func.}}
  & Cache Hit Semantics & Hits match prior stores. Secure stores have authoritative origins. & \cmark & \cmark & \SI{2238}{\second} & Tamarin \\
2 & & Cache Miss Handling & Stores require a prior miss, expiry, or insecure hit, and successful validation of the stored data. & \cmark & \cmark & \SI{195}{\second} & Tamarin \\
3 & & Server Partitioning & Zone-specific cache keys prevent cross-zone reuse. & \cmark & \cmark & \SI{100}{\second} & Tamarin \\
4 & & Atomic Expiration & After expiry deletion, a hit requires a fresh matching store. & \cmark & \cmark & \SI{23.8}{\hour} & Tamarin \\
\cmidrule{1-8}

5 & \multirow{3}{*}{\shortstack[l]{Cache -- \\Consistency}}
  & Mutual Exclusion & Lock release requires a prior acquisition event. Mutual exclusion follows from SAPIC+ lock semantics. & \cmark & \cmark & \SI{99.4}{\second} & Tamarin \\
6 & & State Visibility & Cache hits follow store completion under the model restriction. & \cmark & \cmark & \SI{90.7}{\hour} & Tamarin \\
7 & & Validation Segregation & Validated and unvalidated responses should be stored separately. Otherwise, CD-bit queries can poison normal resolution. & \cmark & \xmark & $<$\SI{1}{\second} & ProVerif \\
\cmidrule{1-8}

8 & \multirow{2}{*}{\shortstack[l]{Cache -- \\Liveness}}
  & Termination & Query/response traces exist, including cache-miss and expired-entry scenarios. This is existential reachability. & \cmark & \cmark & \SI{23.1}{\hour} & Tamarin \\
9 & & Resource Reclamation & A response and release of a modeled lock, and a completed query cycle, are reachable. Global reclamation is not proved. & \cmark & \cmark & \SI{19.3}{\hour} & Tamarin \\
\cmidrule{1-8}

10 & \multirow{4}{*}{\shortstack[l]{Data Auth.\\\& Integrity}}
   & Origin Authentication & Any accepted DNS data originates from its authoritative zone owner with a valid signature. & \cmark & \cmark & \SI{0.6}{\second} & Tamarin \\
11 & & Record Validation & Server-published RRsets have matching zone-signature events. & \cmark & \cmark & \SI{0.1}{\second} & Tamarin \\
12 & & Integrity Preservation & Any adversarial modification to signed RRset data invalidates the signature, ensuring tamper detection. & \cmark & \cmark & \SI{0.1}{\second} & Tamarin \\
13 & & Algorithm Integrity & Algorithm changes must not let on-path attackers bypass validation. & \cmark & \xmark & $<$\SI{1}{\second} & ProVerif \\
\cmidrule{1-8}

14 & Chain of Trust
   & Chain Integrity & DS-DNSKEY links are iteratively verified from root trust anchor to target zone, preventing key substitution. & \cmark & \cmark & \SI{2.1}{\second} & Tamarin \\
\cmidrule{1-8}

15 & \multirow{3}{*}{\shortstack[l]{Denial Auth.\\(NSEC)}}
   & Executability & The protocol successfully produces valid NSEC denial-of-existence responses under honest execution. & \xmark & \cmark & \SI{18.3}{\second} & Tamarin \\
16 & & Domain Secrecy & Domain names remain secret unless explicitly queried. NSEC exposes the next owner name in plaintext, permitting zone enumeration. & \xmark & \xmark & \SI{48.7}{\second} & Tamarin \\
17 & & Result Authentication & NSEC denial responses are cryptographically authenticated via RRSIG with proper label count validation. & \xmark & \cmark & \SI{23.0}{\second} & Tamarin \\
\cmidrule{1-8}

18 & \multirow{3}{*}{\shortstack[l]{Denial Auth.\\(NSEC3)}}
   & Executability & The protocol successfully produces valid NSEC3 denial-of-existence responses under honest execution. & \xmark & \cmark & $<$\SI{1}{\second} & ProVerif \\
19 & & Domain Secrecy & Attacker knowledge of a symbolic name requires a resolver-knowledge event. This check excludes guessing and imposes no event ordering. & \xmark & \cmark & $<$\SI{1}{\second} & ProVerif \\
20 & & Result Authentication & NSEC3 denial responses are cryptographically authenticated via RRSIG with closest encloser proof validation. & \xmark & \cmark & $<$\SI{1}{\second} & ProVerif \\
\cmidrule{1-8}

21 & \multirow{3}{*}{\shortstack[l]{Cross-RFC\\Interactions}}
   & Wildcard Signer Containment & The wildcard owner reconstructed from RRSIG \texttt{Labels} must lie within the signer's zone. Otherwise, child zones poison parent wildcards. & \cmark & \xmark & $<$\SI{1}{\second} & ProVerif \\
22 & & Denial Exclusivity & Modeled NSEC and NSEC3 views should not yield conflicting existence and denial outcomes for the same name. & \cmark & \xmark & \SI{376}{\second} & Tamarin \\
23 & & Validation Preservation & Synthesis requires validated NSEC records. Otherwise, NTA-cached records can be exploited after NTA expiration. & \cmark & \xmark & $<$\SI{1}{\second} & ProVerif \\

\bottomrule
\end{tabular*}

\vspace{0.5em}
\footnotesize
\raggedright
Note: \cmark/\xmark\ in the Verified column indicate whether the property holds. Properties~7, 13, 16, and 21--23 are violated. Section~\ref{subsec:additional_findings} describes P22--P23. Section~\ref{subsec:rfc_survey} gives the cross-RFC survey context.
Results and times are study-reported. Properties~8--9 are reachability checks.
\end{table*}

\subsection{A Motivating Example} %
\label{subsec:compositional_challenges}

Denial-of-existence illustrates the cross-RFC challenges. Figure~\ref{fig:composition_diagram} shows how validation, denial proofs, and auxiliary modes interact through a shared cache.
These are schematic processing steps, not exhaustive RFC-defined state machines. The dashed links mark audited dependencies, not mandated RFC transitions. Aggressive caching can synthesize both negative answers and positive wildcard answers locally when suitable reusable evidence is available~\cite{rfc8198}.

\begin{figure}[ht!]
\centering
\includegraphics[page=2,width=\columnwidth]{figures.pdf}
\caption{
Cross-RFC interactions through a shared cache. Solid arrows show processing and data dependencies, dashed links mark audited interactions, and shading identifies security-sensitive state and data. %
}
\label{fig:composition_diagram}
\Description{Three upper RFC panels feed a shared cache above an aggressive-synthesis panel. The negative-trust-anchor lifecycle shows validation, active bypass, expiration, and resumed validation. The core DNSSEC panel shows owner reconstruction from RRSIG Labels, signature verification, canonical ordering, and NSEC proof checks. The NSEC3 panel shows hashing, closest-encloser, next-closer and wildcard proofs, and an Opt-Out check. The shared cache holds validation status, wildcard RRsets, and denial proofs. Reuse must respect state, zone scope, and remaining lifetime. Dashed edge 1 highlights a wildcard owner reconstructed above the signer and cached as Secure. Edge 2 links NTA bypass to cached validation status. Edge 4 links expiration to possible cache retention. Edge 3 shows an implementation-dependent NSEC-to-NSEC3 fallback. Below, query and cache lookup branch to wildcard plus parent-NSEC evidence, a complete negative proof, or no usable proof. The highlighted P21 path produces a positive A answer with NOERROR and AD=1, conditional on suitable evidence and a vulnerable resolver. The other branches produce NXDOMAIN or NODATA, or query an authority.}
\end{figure}

Verifying cross-RFC interactions introduces challenges distinct from single-RFC verification. We identify three key challenges that prior DNSSEC verification efforts~\cite{kammuller2014verification,eixarch2012formal,chetioui2019formal} did not address. These challenges correspond to the numbered interactions and highlighted state and cache entries in Figure~\ref{fig:composition_diagram}.

\vspace{2pt} \noindent \textbf{Challenge 1: Cross-RFC Under-Specification} (interactions \circled{1} and \circled{4} in Figure~\ref{fig:composition_diagram}).
RFC~4035 \S5.3.2 reconstructs wildcard owners from RRSIG \texttt{Labels}, while RFC~8198 \S5.3 permits reuse for synthesis (interaction~\circled{1}, P21). Neither RFC~4034 \S3.1.3 nor RFC~4035 requires the reconstructed owner to remain within the signer's zone. A child-zone signature can therefore authenticate a parent-zone wildcard that a vulnerable resolver reuses with suitable parent NSEC evidence (Section~\ref{subsubsec:wildcard_containment}).
The 14 model-ready edges instantiate the three failure patterns represented by P21--P23 (Section~\ref{subsubsec:composition_survey}). Single-RFC verification~\cite{kakarla2022scale} may miss these interactions.

\vspace{2pt} \noindent \textbf{Challenge 2: Temporal Security State Tracking} (interactions \circled{2} and \circled{4}: NTA bypass and post-expiry cache retention, P23).
Auxiliary modes like NTA and aggressive caching (Section~\ref{subsec:dns_dnssec}) make the security properties of DNSSEC records context-dependent: a record's trustworthiness depends on the validation state \emph{at the time it was cached}, which may differ from the resolver's current state. DNS-Maude~\cite{liu2023formal} omits this validation state.

\vspace{2pt} \noindent \textbf{Challenge 3: Lack of Support For Ordering Semantics} (interaction \circled{3}: implementation-dependent NSEC$\rightarrow$NSEC3 fallback, P22).
Denial-of-existence creates asymmetric complexity compared to positive authentication. Proving absence requires an ordered space. NSEC orders names, whereas NSEC3 orders their hashes.

For existing names \texttt{a.example} and \texttt{b.example}, NSEC covers the open interval between the names, while NSEC3 covers an interval between their hashes.
Thus, \texttt{c.example} lies beyond the NSEC interval but may hash into the NSEC3 interval.
Both orderings therefore need explicit predicates in the symbolic model.

\subsection{Modeling Approach}
\label{subsec:modeling_approach}

Our model represents three SAPIC+ roles (Section~\ref{subsec:tamarin}). \textit{Clients} initiate queries. \textit{Authoritative nameservers (NSes)} manage zones, sign records, and publish delegation chains. \textit{Recursive resolvers} resolve queries, validate signatures, and cache responses. The resolver combines core DNSSEC (RFC~4033--4035), NTA (RFC~7646), aggressive caching (RFC~8198), and NSEC3 (RFC~5155). Each query requires a cache check that respects validation state, followed by recursion on a miss, signature validation, and denial-of-existence handling.

\vspace{2pt} \noindent \textbf{Messages with Validation Context.}
We extend DNS message models~\cite{rfc1035,rfc4035} with \emph{validation context} metadata. Queries retain $\langle$\texttt{QNAME, QTYPE}$\rangle$, while responses use $\langle$\texttt{RNAME, RTYPE, RDATA, ValState, DType}$\rangle$. The \texttt{ValState} labels $\mathsf{secure}$, $\mathsf{insecure}$, and $\mathsf{bogus}$ distinguish validated, insecure, and failed-validation responses.
The \texttt{DType} field ($\mathsf{nsec} | \mathsf{nsec3} | \mathsf{both}$) indicates which denial mechanism was used, and we use $\mathsf{both}$ to indicate both NSEC and NSEC3 records exist in the same zone.

\vspace{2pt} \noindent \textbf{Two-Dimensional Cache State.}
Prior work like DNS-Maude~\cite{liu2023formal} models TTL semantics probabilistically, but probabilistic TTL models do not capture the core problem we encountered: cached entries may \emph{outlive their validation context}. We implement a \emph{validation-aware cache} that tracks security state alongside freshness. Each cache entry is a tuple:
\[
\mathsf{CacheEntry}(\mathit{rr}, \mathit{fresh}, \mathit{val\_state})
\]
Here $\mathit{fresh} \in \{\mathsf{active},\mathsf{expired}\}$, and $\mathit{val\_state}$ records validation context. The bounded cache stores validated data as $\mathsf{secure}$ and returns $\mathsf{bogus}$ on validation failure. The separate NTA models distinguish validated and unvalidated cache contents. After NTA expiry, \emph{entries may retain their earlier validation status}.

\vspace{2pt} \noindent \textbf{Zone Hierarchy and Configuration.} Our fixed hierarchy has one client and four NSes: a root, a TLD, and two SLDs, following RFC~4035's signed-zone example. Separate models represent NSEC and NSEC3. This scope is consistent with DNS-Maude~\cite{liu2023formal} and exercises the selected cross-RFC interactions. Results depend on the encoded zone contents and denial proofs. Appendix~\ref{sec:append_model} describes the fixed hierarchy, and the artifact contains the zone models.
\section{Formal Analysis via SAPIC+}
\label{sec:design}

\subsection{SAPIC+ Model Architecture}
\label{subsec:model_arch}

We model the three-role DNSSEC architecture from Section~\ref{subsec:modeling_approach} as concurrent SAPIC+ processes communicating over Dolev-Yao channels. 
The artifact contains the SAPIC+ sources and property map (Appendix~\ref{app:open_science}). Models use the fixed hierarchy in Section~\ref{subsec:modeling_approach} and selected zone contents. Appendix~\ref{sec:append_model} details the encoding and assumed interfaces.

\vspace{2pt} \noindent \textbf{Two-Dimensional Cache State.}
The cache tracks freshness and validation state to address Challenge~2 (Temporal Security State Tracking). Expired entries trigger fresh queries. Aggressive synthesis requires an unexpired entry with $\mathit{val\_state} = \mathsf{secure}$. Data cached during an NTA window must pass validation before such reuse. The validation models assume this interface (Section~\ref{sec:evaluation}).

\vspace{2pt} \noindent \textbf{Concurrency via Lock Primitives.}
The bounded cache model uses process replication and entry-level \texttt{lock}/\texttt{unlock} primitives. Query processes acquire a lock before cache access and release it before network communication. SAPIC+'s locking semantics provide mutual exclusion. RFC~4035 \S4.5's atomic cache entry concerns record lifetimes, while entry-level locking is our modeling choice. Properties~5--6 check modeled lock and store ordering.

\vspace{2pt} \noindent \textbf{Chain of Trust Verification.}
We model iterative trust chain validation as a guarded process. Algorithm~\ref{alg:chain-verif} shows the pseudocode. The SAPIC+ encoding uses sequential message exchanges with hash verification at each level.

\begin{algorithm}[h]
\caption{Chain of Trust Verification in SAPIC+}
\label{alg:chain-verif}
\begin{algorithmic}[1]
\Require Trust anchor $\mathsf{DS}_0$
\Ensure Valid chain of trust or abort
\State $i \gets 0$
\While{$i < n$}
    \State Fetch $\mathsf{DNSKEY}_{i+1}$ from zone $i+1$
    \If{$\mathsf{Hash}(\mathsf{DNSKEY}_{i+1}) \neq \mathsf{DS}_i$}
        \State \textbf{abort} \Comment{Chain validation failed}
    \EndIf
    \State Compute $\mathsf{DS}_{i+1}$ from $\mathsf{DNSKEY}_{i+1}$
    \State $i \gets i + 1$
\EndWhile
\State \textbf{return} Valid \Comment{Chain of trust established}
\end{algorithmic}
\end{algorithm}

The algorithm enforces that the resolver proceeds to level $i+1$ only if $\mathsf{Hash}(\mathsf{DNSKEY}_{i+1}) = \mathsf{DS}_i$, preventing key substitution attacks under Dolev-Yao assumptions.

\vspace{2pt} \noindent \textbf{Wildcard Owner Reconstruction via RRSIG \texttt{Labels}.}
RFC~4035 \S5.3.1 lists eight RRSIG validity conditions. Section~5.3.2 reconstructs the canonical owner as $\texttt{*.}\Vert\mathsf{rightmost}(\sigma.\mathsf{labels},\mathit{owner})$ when $\sigma.\mathsf{labels} < |\mathit{owner}|$. The minimal \texttt{rrsig\_underflow} model records conditions~1--5 without independent checks, including the upper bound and signature time validity. Condition~6 uses a restriction over ground bailiwick facts, and condition~7 checks a symbolic signature. Names and label counts are selected constants. Cached-NSEC presence abstracts interval coverage and closest-encloser checks. These choices limit the result to the encoded construction.
The \texttt{Cached\_Wildcard} event records insertion of a reconstructed wildcard at state \emph{Secure}. The \texttt{zone\_boundary\_integrity} query links an authenticated answer to an earlier bailiwick event (Property~21, Section~\ref{subsubsec:wildcard_containment}). The reported ProVerif counterexample uses $\mathit{wc}=\texttt{*.test.}$ and $\mathit{signer}=\texttt{attacker.test.}$. A companion model adds $|\mathit{signer}| \leq \sigma.\mathsf{labels}$ and excludes this construction. The NTA models separately record bypass, cache insertion, expiry, and reuse.

\subsection{Baseline Security Properties}
\label{subsec:baseline_props}

Before analyzing RFC interactions, we check baseline properties of the selected SAPIC+ models. Properties~1--6, 8--12, 14--15, and 17--20 hold under their respective abstractions. Properties~7, 13, and 16 are violated (Table~\ref{tab:verification-results}). These results apply to the modeled behaviors and assumptions, rather than to all DNSSEC executions.

\subsubsection{Cache Correctness (Properties 1--6, 8--9)}

We verify eight properties ensuring the resolver's internal cache maintains correctness under concurrent operations.

\vspace{2pt} \noindent \textbf{Core Functionalities (Properties 1--4).}
Property~1 links each cache hit to a prior store of the same data. Secure stores also require a prior authoritative response. Grouping cache keys, data, and validation metadata as $x$, we write:
\[
\begin{aligned}
\forall x,t:\ &\mathsf{Hit}(x)@t \implies \exists u<t:\ \mathsf{Store}(x)@u
\end{aligned}
\]
Property~2 requires a prior miss, expiry, or insecure hit before a store, plus successful validation of the stored data. These backward-looking checks establish prerequisites for stores without guaranteeing completion of every miss.

Zone-specific keys prevent cross-zone reuse (Property~3). Expiry deletes entries before refresh. P4 requires a store between deletion and a later hit. We suppress record arguments:
\[
\begin{aligned}
\forall Q,t_d,t_h:&\ \mathsf{DeleteExpired}(Q)@t_d \land \mathsf{Hit}(Q)@t_h \land t_d<t_h\\
&\implies \exists t_s: \mathsf{Store}(Q)@t_s \land t_d<t_s<t_h
\end{aligned}
\]

\vspace{2pt} \noindent \textbf{Concurrency Safety (Properties 5--6).}
SAPIC+'s lock semantics provide mutual exclusion. For entry $e$, the semantic requirement is:
\[
\square(\forall T_1, T_2: T_1 \neq T_2 \implies \neg(\mathsf{Holds}(T_1, e) \land \mathsf{Holds}(T_2, e)))
\]
The explicit P5 query requires a lock-acquisition event before release. P6 requires store completion before a cache hit, under the model's store-completion restriction. These checks concern event ordering within the bounded model.

\vspace{2pt} \noindent \textbf{Progress Reachability (Properties 8--9).}
The bounded cache model includes existential traces in which queries receive responses, including cache-miss and expired-entry scenarios (Property~8). Property~9 checks reachability of a response and release of a modeled lock, and of a completed query cycle. These checks establish that progress is possible. They do not prove that every query terminates, impose an order between the response and unlock, or establish global lock reclamation under all schedules.

\subsubsection{Data Origin Authentication (Properties 10--12)}

DNSSEC ensures data origin authentication by binding digital signatures (RRSIG) to RRsets, validated through zone keys (DNSKEY) and a hierarchical chain of trust rooted in trust anchors (DS records). We verify the authentication guarantee:
\[
\begin{aligned}
\forall D:\ &\mathsf{Accepted}(D) \implies \exists Z,\sigma:\ \mathsf{Owns}(Z,D) \land \mathsf{IssuedSig}(Z,D,\sigma)
\end{aligned}
\]

Here $\mathsf{Owns}(Z,D)$ denotes zone $Z$'s authority over $D$, and $\mathsf{IssuedSig}$ records its signature. The model assumes unforgeable signatures and trusted DS-DNSKEY links. These checks establish symbolic correspondence, without a computational reduction.

Property~11 links server-published RRsets to zone-signature events. For a valid signature $\sigma$ on $D$, P12 rejects altered symbolic data $D' \neq D$:
\[
\mathsf{Verify}(K, \mathsf{RRData}(D'), \sigma) = \mathsf{false}
\]

\subsubsection{Chain of Trust Verification (Property 14)}

The hierarchical trust structure is established through DS records binding child DNSKEYs to parent zones. Validation requires:
\[
\begin{aligned}
\forall i:\ &\mathsf{Hash}(\mathsf{DNSKEY}_{i+1}) = \mathsf{DS}_i \land\ \mathsf{TrustAnchor}(\mathsf{DS}_0)
\end{aligned}
\]
where $\mathsf{TrustAnchor}(\mathsf{DS}_0)$ represents the root zone's pre-configured trust anchor. The resolver iteratively verifies each DS-DNSKEY link starting from the root anchor down to the target zone (see Algorithm~\ref{alg:chain-verif} in Section~\ref{subsec:model_arch}).

Under Dolev-Yao, an adversary cannot inject a fraudulent DNSKEY at level $i$ without possessing a preimage of $\mathsf{DS}_{i-1}$. Since hash functions are modeled as one-way, key substitution is infeasible. The property \textbf{Chain Integrity} holds: \textit{given valid DS-DNSKEY links and a trusted $\mathsf{DS}_0$, a Dolev-Yao adversary cannot make the resolver accept a forged DNSKEY.}

\subsubsection{Domain Name Secrecy (Properties 16, 19)}

NSEC3 replaces plaintext owner names with hashes. P19 concerns fresh symbolic names, excluding structured-name guessing. Let $\mathsf{SrvName}$ denote server domain hosting and $\mathsf{RslvKnow}$ denote resolver knowledge of a name. The retained query states:
\[
\begin{aligned}
& \forall s,d,t,u:\ \mathsf{SrvName}(s,d)@t \land \mathsf{K}(d)@u\\
& \quad \implies \exists r,v:\ \mathsf{RslvKnow}(r,d)@v
\end{aligned}
\]
Here $\mathsf{K}(d)$ is attacker knowledge. The query imposes no ordering between $u$ and $v$ and gives no computational bound on offline enumeration. RFC~5155 acknowledges dictionary attacks~\cite[\S12.1.1]{rfc5155}.

Property~16 (NSEC secrecy) is \emph{violated} as expected: NSEC records expose next domain names in plaintext, enabling zone walking. Property~19 holds only in our symbolic abstraction. It does not imply that NSEC3 prevents practical enumeration, as illustrated by prior work on structured IPv6 reverse zones~\cite{borgolte2018enumerating} (Section~\ref{sec:discussion}).

\subsection{Authentication of Denial of Existence}
\label{subsec:auth_of_doe}

The NSEC/NSEC3 protocols enforce authenticated denial of existence through cryptographic proof chaining and strict validation of RRSIG scoping. We formalize these guarantees through the following theorem. Let $\mathsf{RcvResult}$ and $\mathsf{SndResp}$ abbreviate resolver reception and server transmission events:
\[
\begin{aligned}
& \forall s, r, q, m, \#i:\ \mathsf{RcvResult}(s, r, q, m)@\#i \\
& \quad \implies \exists \#j, \sigma, n, n':\ \mathsf{SndResp}(s, r, q, m)@\#j \land \#j < \#i \\
& \quad \land\ m = \{\mathsf{NSEC}(n, n') \lor \mathsf{NSEC3}(h(n), h(n'))\}_{\sigma} \\
& \quad \land\ \mathsf{ValidCover}(q, n, n') \land \mathsf{ValidLabel}(\sigma, q)
\end{aligned}
\]

$\mathsf{ValidCover}(q, n, n')$ verifies $q \notin \mathsf{zone}$ with $n < q < n'$ in the zone's canonical order (or hashed order for NSEC3). $\mathsf{ValidLabel}(\sigma, q)$ ensures the RRSIG's label count matches the queried name's depth, excluding wildcards, \emph{and} that $\sigma.\mathsf{labels} \geq |\mathsf{signer}(\sigma)|$, i.e., the reconstructed wildcard owner is a subdomain of the signer's name. The latter constraint is \emph{not} mandated by RFC~4034/4035. Our model identifies this missing invariant and a counterexample when it is omitted (Section~\ref{subsec:aggressive_caching}).

This theorem establishes three critical properties for any accepted denial-of-existence message: (1)~cryptographic authenticity through server signatures, (2)~logical proof of non-existence via adjacent zone entries, and (3)~correct signature scope enforcement through label count validation. Temporal consistency is guaranteed through the ordering constraint $\#j < \#i$ between server transmission and resolver acceptance.

\vspace{2pt} \noindent \textbf{NSEC3 Closest Encloser Proofs.}
NSEC3 denial proofs require three records~\cite[\S 8.4--8.5]{rfc5155}: one matching the hash of the closest encloser $\mathit{ce}$ (the longest existing ancestor of $q$), one covering the hash of the next closer name $\mathsf{NC}(q,\mathit{ce})$, and one covering the hash of the wildcard $\texttt{*.}\mathit{ce}$. Our predicate encodes this directly as $\mathsf{VCov}_{\mathsf{N3}}(q) \iff \exists\,\mathit{ce}.\ \mathsf{CE}(q,\mathit{ce}) \land \mathsf{Cov}(h(\mathsf{NC}(q,\mathit{ce}))) \land \mathsf{Cov}(h(\texttt{*.}\mathit{ce}))$, ensuring semantic consistency (Challenge~3, Section~\ref{subsec:compositional_challenges}): NSEC3's hash-based ordering must be interpreted relative to the queried name's position in the zone hierarchy.

\vspace{2pt} \noindent \textbf{Wildcard Handling.}
For wildcard denial, $\mathsf{ValidCover}$ excludes names with an applicable wildcard source. RFC~4592 \S3.3.1 defines that source as $\texttt{*.}$ followed by the closest encloser~\cite{rfc4592}. It selects one candidate, without searching another wildcard if that source is absent. We use this rule for NSEC and NSEC3 alongside the signer-containment condition in $\mathsf{ValidLabel}$.

Properties~15 and 18 establish executability, while Properties~17 and 20 establish result authentication in the separate NSEC and NSEC3 models (Table~\ref{tab:verification-results}). These authentication checks do not establish resistance to zone enumeration.

\subsection{Cross-RFC Properties}
\label{subsec:aggressive_caching}

RFC~8198 ``aggressive caching'' specifies that a resolver can synthesize denial-of-existence responses with cached NSEC/NSEC3 records, as described in Section~\ref{subsec:dns_dnssec}.
While this optimization improves performance, it composes with the wildcard-reconstruction logic of RFC~4034/4035, the validation bypass of RFC~7646, and the dual ordering semantics of RFC~5155. Our formal analysis systematically explores whether RFC~8198's implicit assumptions hold under each of these compositions.

\subsubsection{RFC~4034 $\times$ RFC~4035 $\times$ RFC~8198: Wildcard Zone-Cut Containment}
\label{subsubsec:wildcard_containment}

RFC~8198 \S5.3 permits a resolver to synthesize a positive answer from a cached wildcard RRset plus an NSEC record proving no closer match exists. The wildcard RRset, in turn, is cached after RFC~4035 \S5.3.2 validation: when a response's RRSIG has $\sigma.\mathsf{labels} < |q|$, the validator reconstructs the canonical owner as ``\texttt{*.}'' prepended to the rightmost $\sigma.\mathsf{labels}$ labels of $q$ and verifies the signature over that form. RFC~4034 \S3.1.3 upper-bounds $\sigma.\mathsf{labels} \leq |q|$ but neither RFC lower-bounds it against the signer's depth. Consequently, a child zone $\mathsf{signer}(\sigma) = \texttt{attacker.test.}$ can set $\sigma.\mathsf{labels} = 1$ and the reconstructed owner becomes \texttt{*.test.}, a name in the \emph{parent} zone.

\vspace{2pt} \noindent \textbf{Property 21: Wildcard Signer Containment.}
A wildcard RRset used for RFC~8198 synthesis SHALL have an owner within its signer's zone. We write $\sqsubseteq$ for the subdomain-or-equal relation:
\[
\begin{aligned}
\forall \mathit{rr}, \sigma, t:\ &\mathsf{CachedWildcard}(\mathit{rr}, \sigma, t) \land \mathsf{UsedForSynthesis}(\mathit{rr}, t) \\
&\implies \mathsf{WildcardOwner}(\mathit{rr}) \sqsubseteq \mathsf{Signer}(\sigma)
\end{aligned}
\]
The formula states the intended containment requirement. The executable query requires an \texttt{AD\_Answer\_Signed\_By} event to have an earlier matching \texttt{InBailiwick} event. In the reported counterexample, a child-zone signature causes reuse of a parent-zone wildcard outside the child's authority. Section~\ref{subsec:pitfalls} describes the corresponding packet scenario.

\subsubsection{Further Cross-RFC Properties}
\label{subsubsec:further_composition}

Two further cross-RFC properties concern false denial. Section~\ref{subsec:additional_findings} describes their attack conditions and resolver outcomes.

\vspace{2pt} \noindent \textbf{Property 22 (RFC~4034 $\times$ RFC~5155 $\times$ RFC~8198): Denial Exclusivity.}
The artifact's \texttt{denial\_exclusivity} lemma forbids conflicting existence and denial outcomes for the same modeled name:
\[
\begin{aligned}
\forall r_1,r_2,q,t_1,t_2:\ &\mathsf{Exists}(r_1,q,t_1) \land\mathsf{Denies}(r_2,q,t_2) \implies \bot
\end{aligned}
\]
Here \(\mathsf{Exists}\) and \(\mathsf{Denies}\) abbreviate the model's resolver-result events. The counterexample uses static, inconsistent NSEC/NSEC3 views. Coexistence alone is not proof of false denial or of a one-denial-type-per-cache invariant. Exploitation requires usable signed proofs and a vulnerable resolver path.

\vspace{2pt} \noindent \textbf{Property 23 (RFC~7646 $\times$ RFC~8198): Validation Preservation.}
\[
\begin{aligned}
\forall \mathsf{nsec}, t_1, t_2:\ &\mathsf{Cached}(\mathsf{nsec}, t_1) \land \mathsf{NTA\_Active}(\mathsf{zone}(\mathsf{nsec}), t_1) \\
&\land\ t_2 > \mathsf{NTA\_Expires}(\mathsf{zone}(\mathsf{nsec})) \\
&\implies \neg\mathsf{UsedForSynthesis}(\mathsf{nsec}, t_2)
\end{aligned}
\]
Unvalidated NSEC records must not support synthesis after NTA expiry. The SAPIC+ model contains an ordered expiry-attack witness. Its ProVerif companion checks event co-occurrence, which alone does not establish this ordering. Section~\ref{subsec:additional_findings} states the cache-lifetime and validation conditions needed for the attack.

\subsubsection{Cross-RFC Review of Properties 21--23}
\label{subsubsec:composition_survey}

We now revisit the 52 cross-RFC edges identified in Section~\ref{subsec:rfc_survey} to assess how broadly the failure patterns exposed by Properties~21--23 recur across the DNSSEC specifications. Of these edges, ten are already captured by Properties~1--23, fifteen concern signaling, telemetry, or registry-side behavior outside our scope, and thirteen require model extensions not developed in this work (E1--E5, Section~\ref{sec:discussion}). The remaining fourteen can be analyzed using our existing model. Each instantiates one of three recurring failure patterns exemplified by P21--P23:

\vspace{2pt} \noindent\ding{182} \textbf{Zone-cut containment} (anchor: P21). A denial or wildcard record's scope must not extend across the signer's zone cut. RFC~6840 \S4.1 states this for ancestor-side NSEC/NSEC3. RFC~4470 \S3 and RFC~9824 \S3 impose narrower coverage through $\varepsilon$-width ranges. RFC~6840 \S5.10 constrains opt-out NSEC3. P21 concerns RFC~4035 \S5.3.2's missing wildcard signer-depth bound.

\vspace{2pt} \noindent\ding{183} \textbf{Denial-mechanism exclusivity} (anchor: P22). A resolver's synthesis logic must operate over one consistent ordering semantics per zone per cache epoch. RFC~9824 \S3 (compact DoE sentinel vs.\ range NSEC), RFC~9077 \S3.4 (TTL-epoch alignment), and RFC~6840 \S4.3 (CNAME-bit vs.\ NODATA bitmap) each constrain which cached denial records may be combined. Property~22 is the instance where NSEC and NSEC3 orderings collide.

\vspace{2pt} \noindent\ding{184} \textbf{Validation-state preservation} (anchor: P23). Only records at RFC~4035 state \emph{Secure} may seed synthesis or set the AD bit. RFC~9276 \S3.2 (iteration-downgraded NSEC3), RFC~8767 \S4 (serve-stale data), RFC~8020 \S2 (NXDOMAIN-cut), and RFC~8624 \S3.3 (deprecated algorithms) each move records out of \emph{Secure} and implicitly rely on synthesis honoring that demotion. Property~23 is the instance where RFC~7646 demotes a subtree to \emph{Insecure} but RFC~8198 is silent on the cached-record consequence.

P21--P23 are therefore not isolated: they are the RFC-8198-adjacent instances of patterns that recur across the dnsext/dnsop corpus. 
Appendix~\ref{app:rfc_survey} retains supporting edge listings and pattern assignments. Model-ready edges remain candidates for further analysis, rather than additional verified results.
\section{Evaluation Results}
\label{sec:evaluation}

Our experimental setup used the SAPIC+ Docker environment from the original SAPIC+~\cite{cheval2022sapic+}. We ran all evaluations on a GCP c4d-highmem-384 instance (384 vCPUs, 3,024 GB RAM). Our SAPIC+ model contains 3,928 lines.

Table~\ref{tab:verification-results} in Section~\ref{sec:formal_models} summarizes all 23 properties and their verification results.
Models are implemented in SAPIC+ and translated to ProVerif~\cite{blanchet2018proverif} (v2.04) and Tamarin~\cite{meier2013tamarin} (v1.10.0). The artifact retains SAPIC+ sources and four ProVerif companions. Its property-to-lemma map records their coverage and limitations. The artifact also includes reachability checks for cache initialization, authoritative responses, secure cache hits, and NTA bypass. Honest-signing-key secrecy is a separate safety check. These auxiliary checks are not additional paper properties.
We use Tamarin for stateful cache checks (Properties~1--6, 8--9), with Properties~8--9 encoded as reachability. Property~7 uses the separate ProVerif CD-bit model. Verification times range from seconds to hours. ProVerif is used for NSEC3 denial properties (Properties~18--20) because closest encloser proofs expand the search space beyond Tamarin's tractability. For our novel findings, Property~22 uses Tamarin for a precise counterexample trace. Properties~21 and~23 use ProVerif, the latter because NTA temporal state combined with the cache model exceeds Tamarin's capacity. Both are authored in SAPIC+/Tamarin and exported to ProVerif for automated falsification, because Tamarin's automated checker does not scale to these models. %
A ProVerif ``cannot be proved'' result alone does not establish an attack. We interpret candidate traces against the modeled transitions and the implementation experiments. The P21 packet sequence is described below.

For denial-of-existence authentication properties (Properties 15--20),
we analyze separate validation/denial models using an assumed cache interface. Entries must be unexpired, preserve validation state, and support mutually exclusive access, following assume-guarantee reasoning~\cite{abadi1993composing}. We provide no machine-checked composition theorem. Table~\ref{tab:verification-results} reports P7, P13, and P16 violations, so these checks do not prove all baseline properties under composition.

\begin{table*}[t]
\centering
\caption{Summary of findings. The first three rows concern Properties~21--23 (Section~\ref{subsec:aggressive_caching}). The last three are known attacks rediscovered by the model (Section~\ref{subsec:rediscovery}).}
\label{tab:findings_summary}
\footnotesize
\setlength{\tabcolsep}{4pt}
\renewcommand{\arraystretch}{1.15}
\begin{tabularx}{\textwidth}{@{}p{0.145\textwidth}p{0.12\textwidth}Xp{0.15\textwidth}p{0.095\textwidth}@{}}
\toprule
\textbf{Finding} & \textbf{RFCs} & \textbf{Vulnerability Description} & \textbf{Adversary} & \textbf{Impact} \\
\midrule
\multicolumn{5}{@{}l}{\textit{Novel cross-RFC vulnerabilities}} \\
\addlinespace[2pt]
\textbf{Cross-Zone Wildcard Cache Poisoning} (P21) &
4034 $\times$ 4035 $\times$ 8198 &
RFC~4035 \S5.3.2 lacks a lower bound on RRSIG \texttt{Labels}. A child signs \texttt{*.parent.}, which an affected resolver accepts as \emph{Secure}. Vulnerable RFC~8198 paths forge AD answers for siblings covered by parent NSEC. &
Off-path signed child under an NSEC parent &
Integrity (AD-bit forgery) \\
\addlinespace[3pt]
\textbf{NSEC/NSEC3 Coexist.} (P22) &
4034 $\times$ 5155 $\times$ 8198 &
A resolver reuses inconsistent cached NSEC/NSEC3 proofs during fallback and returns NXDOMAIN for an existing name. Coexistence alone does not establish the attack. &
Attacker supplying usable inconsistent proofs &
DoS (false NXDOMAIN) \\
\addlinespace[3pt]
\textbf{NTA Cache Poisoning} (P23) &
7646 $\times$ 8198 &
Unvalidated denial records survive NTA expiry and are reused without revalidation in aggressive synthesis or ordinary negative caching. &
On-path during NTA &
DoS (false NXDOMAIN) \\
\midrule
\multicolumn{5}{@{}l}{\textit{Known attacks rediscovered (model validation)}} \\
\addlinespace[2pt]
\textbf{NSEC Zone Walk}~\cite{wala2024zone} (P16) &
4034 &
NSEC exposes the next owner name in plaintext. Walking the chain enumerates the zone. P19 holds only under the symbolic-name abstraction. &
Any DNS querier &
Privacy (zone enum.) \\
\addlinespace[3pt]
\textbf{Algo. Downgrade}~\cite{heftrig2023downgrading} (P13) &
4034 $\times$ 6840 &
Rewriting the RRSIG algorithm to an unsupported value downgrades the modeled zone to \emph{Insecure}, allowing unsigned data. &
On-path &
Integrity bypass \\
\addlinespace[3pt]
\textbf{CD-Bit Cache Poisoning}~\cite{zhang2025your} (P7) &
4035 &
A CD$=$1 query stores an unvalidated record in a unified cache. Later CD$=$0 queries hit it and receive SERVFAIL. &
Off-path CD$=$1 client &
DoS (SERVFAIL) \\
\bottomrule
\end{tabularx}
\end{table*}

Table~\ref{tab:findings_summary} summarizes the six security findings: three novel RFC-composition vulnerabilities (Properties~21--23) and three previously known attacks our model rediscovers as a consistency check. %

\subsection{Cross-Zone Wildcard Cache Poisoning}
\label{subsec:pitfalls}

The formal analysis reports three cross-RFC violations (Properties~21--23, Table~\ref{tab:verification-results}). We interpret the modeled scenarios through source inspection and controlled resolver tests. Historical prover logs are absent from the retained artifact, limiting independent checks of discovery order and trace correspondence. Property~21 concerns a signed child zone poisoning a parent-zone wildcard and producing forged A answers with AD set. Section~\ref{subsec:additional_findings} describes the two availability findings. Appendices~\ref{sec:code_analysis} and~\ref{app:additional_vulns} retain historical source-code analysis and detailed workflows.

\vspace{2pt} \noindent \textbf{Attacker Goals and Capabilities.}
For P21, a \textit{delegated zone controller} uses its own signed child zone under an NSEC-signed parent. The attacker needs its own signing key and a vulnerable downstream synthesis path, without an on-path position or operator misconfiguration. Clients of the poisoned resolver can receive attacker-chosen A records for some sibling names with AD set. Suitable parent denial evidence limits the affected names.

\vspace{2pt} \noindent \textbf{Root Cause.}
RFC~4034 \S3.1.3 requires only that the RRSIG \texttt{Labels} field not exceed the owner-name label count. RFC~4035 \S5.3.2 instructs the validator, when \texttt{Labels} is strictly less than the owner label count, to reconstruct the canonical owner as ``\texttt{*.}'' prepended to the rightmost \texttt{Labels} labels of the query name and to verify the signature over that form. Neither RFC requires the reconstructed owner to lie within the signer's zone. A child zone \texttt{attacker.test.} can therefore sign an A RRset for \texttt{www.attacker.test.} with \texttt{Labels} $=$ 1. The validator reconstructs the canonical owner as \texttt{*.test.}, one label \emph{above} the zone cut, and the signature verifies because the attacker computed it over exactly that canonical form with its own key. RFC~8198 \S5.3 then permits caching \texttt{*.test.} as \emph{Secure} and synthesizing positive answers.

\begin{figure}[t]
\centering
\includegraphics[page=3,width=\columnwidth]{figures.pdf}
\caption{Cross-zone wildcard cache poisoning via RRSIG \texttt{Labels} underflow and vulnerable synthesis (P21).
}
\label{fig:rrsig_labels_attack}
\Description{Two-phase swimlane diagram with client, resolver, attacker nameserver, and parent nameserver lifelines and thirteen numbered steps. The attacker controls signed child attacker.test under NSEC-signed parent test, without an on-path position. In cache priming, Q1 requests www.attacker.test with type A. The child returns 6.6.6.6 with an RRSIG signed by attacker.test and Labels set to 1. A resolver callout shows chain validation, reconstruction of *.test above the signer, and caching its A record as Secure. Q2 requests nonexistent bystander.test. The parent returns NXDOMAIN and a signed NSEC from attacker.test to zz.test. A cache callout states that it covers nope.test. In Phase 2, Q3 requests nope.test with type A. Three compact resolver steps combine the NSEC's no-closer-match evidence with the cached wildcard to synthesize 6.6.6.6. The response has NOERROR and the authenticated-data flag set. A cross on the parent lifeline indicates no upstream query. The final callout limits the forgery to some NSEC-covered siblings and the usable cache lifetime, on the illustrated vulnerable synthesis path.}
\end{figure}

\vspace{2pt} \noindent \textbf{Attack Scenario.}
Figure~\ref{fig:rrsig_labels_attack} shows three queries. \textbf{Q1} (poison): the resolver requests \texttt{www.attacker.test./A}. The child returns \texttt{A 6.6.6.6} with an RRSIG signed by \texttt{attacker.test.} and \texttt{Labels}$=$1. The resolver validates the child's chain, reconstructs \texttt{*.test.}, and caches its A RRset as \emph{Secure}. \textbf{Q2} (seed): a selected nonexistent sibling query obtains genuine parent NSEC evidence covering the intended victim, \texttt{nope.test.}. \textbf{Q3} (exploit): a client requests \texttt{nope.test./A}. The vulnerable resolver combines the cached wildcard with suitable no-closer-match evidence and returns \texttt{NOERROR AD 6.6.6.6} without an upstream query. Persistence depends on usable wildcard and denial-proof lifetimes, signature checks, and cache policy.

\vspace{2pt} \noindent \textbf{Formal Verification.}
Property~21 states the signer-containment requirement (Section~\ref{subsubsec:wildcard_containment}). The reported ProVerif counterexample uses a child-zone signature and a cached parent-zone wildcard outside that signer's authority. Q1--Q3 describes the corresponding implementation scenario. Adding $\sigma.\mathsf{labels} \geq |\mathsf{signer}(\sigma)|$ excludes the modeled construction. The result concerns the stated abstraction rather than complete implementation remediation.

\begin{table*}[t]
\centering
\caption{Aggressive-caching cross-RFC vulnerabilities across six resolvers. Results refer to the listed tested versions. \textbf{Labels} concerns wildcard processing and synthesis. \textbf{Coexist} and \textbf{NTA} concern availability. The BIND mixed-response crash does not require an honest mixed deployment. NTA findings require validation bypass.}
\label{tab:aggressive_vuln}
\small
\setlength{\tabcolsep}{3.5pt}
\begin{tabular}{llccccclll}
\toprule
\multirow{2}{*}{\textbf{DNS Software}} & \multirow{2}{*}{\textbf{Version}} & \textbf{RFC 8198} & \textbf{RFC 7646} & \multicolumn{3}{c}{\textbf{Vulnerable?}} & \multicolumn{3}{c}{\textbf{Root Cause}} \\
\cmidrule(lr){5-7} \cmidrule(lr){8-10}
 & & \textbf{(Aggr.)} & \textbf{(NTA)} & \textbf{Labels} & \textbf{Coexist} & \textbf{NTA} & \multicolumn{1}{c}{\textbf{Labels}} & \multicolumn{1}{c}{\textbf{Coexist}} & \multicolumn{1}{c}{\textbf{NTA}} \\
\midrule
\textbf{Unbound}~\cite{Unbound}       & 1.24.2  & \cmark$^\dagger$ & \cmark & \CIRCLE$^{\dagger\dagger}$ & \Circle$^\ddagger$   & \Circle     & No lower bound & No NSEC3 synthesis    & Requires Secure state \\
\textbf{Knot}~\cite{KnotDNS}          & 6.2.0   & \cmark           & \cmark & \CIRCLE & \CIRCLE              & \Circle     & No lower bound & Two-slot fallback     & Rank blocks NTA records \\
\textbf{BIND}~\cite{BIND9}            & 9.20.22 & NSEC$^*$         & \cmark & \CIRCLE & \CIRCLE$^{\ddagger\ddagger}$   & \LEFTcircle & No lower bound & Mixed-NSEC crash    & No flush on NTA expiry \\
\textbf{PowerDNS}~\cite{PowerDNS}     & 5.4.0   & \cmark           & \cmark & \CIRCLE & \LEFTcircle$^\S$     & \CIRCLE     & No lower bound & Sequential fallback   & No validation check \\
\textbf{Technitium}~\cite{TechnitiumDNS} & 14.3.0 & \xmark         & \cmark & \LEFTcircle$^{\P\P}$ & \Circle              & \CIRCLE     & Phase 1 only & No aggr. caching & No cache isolation \\
\textbf{Hickory}~\cite{HickoryDNS}    & 0.26.0  & \xmark           & \xmark & \LEFTcircle$^{\P\P}$ & \LEFTcircle$^\P$     & --     & Phase 1 only & Fallback to NSEC3     & No NTA implementation \\
\bottomrule
\end{tabular}
\footnotesize
\flushleft
Symbols: \CIRCLE~vulnerable, \Circle~not vulnerable, \LEFTcircle~partial, -- not implemented.
$^\dagger$NSEC3 for DS queries only.
$^{\dagger\dagger}$Basic AD=1 trace blocked.
$^*$NSEC-only synthesis.
$^\ddagger$No general NSEC3 synthesis.
$^\S$Transition-dependent.
$^\P$Non-validating mode.
$^{\ddagger\ddagger}$Mixed-response crash.
$^{\P\P}$Phase~1 only, no AD=1 forgery.
Test conditions appear in Sections~\ref{subsec:pitfalls}--\ref{subsec:additional_findings}.
\end{table*}

\vspace{2pt} \noindent \textbf{Testing Resolver Software.}
Table~\ref{tab:aggressive_vuln} compares six resolvers. The tested BIND, Unbound, Knot Resolver, and PowerDNS versions share the missing signer-depth bound, but their serving behavior differs. Unbound accepts the malformed wildcard but revalidates synthesized answers, blocking the basic NSEC Q1--Q3 forgery with SERVFAIL. Its separate unsupported-algorithm DS construction returns forged data as \emph{Insecure} (AD=0), as confirmed by NLnet Labs~\cite{unbound2026labels}. The four implementations therefore do not demonstrate the same authenticated-forgery outcome. Technitium and Hickory lack this RFC~8198 synthesis phase.

\vspace{2pt} \noindent \textbf{Vendor Response.}
Following coordinated disclosure, patches for cross-zone wildcard cache poisoning have shipped in all four impacted resolvers: BIND~\cite{isc2026labels}, Unbound~\cite{unbound2026labels}, Knot Resolver~\cite{knot2026dnssec}, and PowerDNS Recursor~\cite{powerdns2026labels}. Table~\ref{tab:aggressive_vuln} lists tested pre-patch versions.

The missing signer-depth bound is shared by the four tested implementations. Their outcomes on the additional findings differ: Unbound and Knot block NTA-cached material using validation-state checks, while Knot is affected by the coexistence case. The experiments establish behavior for the tested versions, not every possible RFC-conformant implementation.

\vspace{2pt} \noindent \textbf{Impact and Exposure.}
The integrity failure is an AD-bit forgery: a client that relies on the resolver's authenticated-data signal accepts attacker-chosen data as DNSSEC-validated. A registrant under an NSEC-signed parent can poison some sibling names when the parent supplies a suitable covering NSEC and the downstream resolver follows a vulnerable wildcard-synthesis path. The demonstrated trace does not overwrite arbitrary existing parent-zone RRsets.
A focused scan on May~1, 2026 found 143,186 open recursive resolvers. Of these, 62,618 accepted a correctly signed test name and rejected a mis-signed sibling. We observed aggressive NSEC synthesis at 3,742 of these validators (6.0\%). These tests establish synthesis support rather than successful exploitation. The earlier 2.24-million-address inventory is a separate cohort described in Appendix~\ref{subsec:resolver_scan}.

Our September~9, 2026 census queried all 1,438 TLDs from IANA's version 2026090900 list~\cite{iana-tld-list}. We used fresh names through two independent recursive services, filtered proofs by TLD signer, and checked root DS records and authoritative responses. Of 53 NSEC-serving TLDs, 52 had a root DS and authenticated responses from both services. The exception, \texttt{.xn--fzc2c9e2c}, lacked a root DS. Among the 52, 51 also returned an authoritative covering NSEC (3.55\% of all TLDs). \texttt{.gov} instead returned compact denial with NXNAME/TYPE128~\cite{rfc9824}. These dated observations establish parent-side prerequisites. Registration rules, target-specific proofs, and resolver behavior limit exploitation. Appendix~\ref{app:open_science} lists the 52 authenticated NSEC-serving TLDs.

The Q1--Q3 trace requires a parent NSEC for Q2 and does not apply unchanged to NSEC3-only parents. NSEC3 synthesis requires zone-scoped closest-encloser proofs. BIND supports aggressive synthesis only with NSEC. Unbound uses NSEC3 for DS queries, not general aggressive synthesis. Our census observes NSEC3 at 1,309 TLDs (91.03\%). This limits the demonstrated trace, without proving NSEC3 configurations immune to related flaws.

The tested Technitium and Hickory versions do not implement RFC~8198 aggressive synthesis, so Phase~2 of the P21 attack does not succeed on them. Their Phase~1 \texttt{Labels} acceptance and the separate P22/P23 test outcomes are distinguished in Table~\ref{tab:aggressive_vuln}.

\subsection{Additional Cross-RFC Findings}
\label{subsec:additional_findings}

Properties~22 and~23 concern false denial under specific cache conditions. Table~\ref{tab:aggressive_vuln} distinguishes these outcomes from the separately reported BIND mixed-response crash.

\vspace{2pt} \noindent \textbf{NSEC/NSEC3 Coexistence} (Property~22, RFC~5155 $\times$ RFC~8198): a resolver caches NSEC and NSEC3 proofs from inconsistent signed views. If NSEC misses a query, an affected implementation tries cached NSEC3 evidence. False denial requires usable inconsistent proofs or a proof-processing flaw. Ordinary coexistence during an RFC~5155 transition alone is insufficient. Our tests produce false denial in Knot. PowerDNS's outcome depends on asymmetric zone-type transitions, while Hickory's default non-validating mode bypasses mixed-record rejection. Unbound and Technitium lack this synthesis path. In BIND~9.20.22, attacker-served mixed NSEC/NSEC3 input triggers a \texttt{RUNTIME\_CHECK} abort under the default configuration, without requiring honest mixed zones. The model and tests do not measure deployment prevalence.

\vspace{2pt} \noindent \textbf{NTA Cache Poisoning} (Property~23, RFC~7646 $\times$ RFC~8198): Figure~\ref{fig:nta_attack} illustrates the attack. An on-path attacker injects unvalidated denial records during an operator-enabled NTA window. If the records survive expiry and are reused without validation, covered existing names can receive false denial. PowerDNS and Technitium tests show cache-persistence issues, with Technitium using ordinary negative caching. BIND's explicit NTA removal clears the cache, while timer expiry differs. Unbound and Knot check validation state. Hickory has no NTA implementation, so its non-validating-mode test does not establish an expiry attack. Cloudflare's May~2026 \texttt{.de} validation override~\cite{cloudflare2026detld} documents operational bypass without establishing P23 exploitation.

\begin{figure}[htbp]
\centering
\includegraphics[page=4,width=\columnwidth]{figures.pdf}
\caption{Conditional NTA cache poisoning (P23). During validation bypass, an on-path attacker injects unvalidated NSEC evidence. False denial after NTA expiry requires that evidence to remain usable and be reused without validation.}
\label{fig:nta_attack}
\Description{Two-phase swimlane diagram with client, resolver, on-path attacker, and authoritative nameserver lifelines and ten numbered steps. A note makes all names relative to zone.com. During Phase 1, an administrator enables an NTA for this zone. The client requests aa with type A. The query passes through the attacker to the authority. The reply contains NSEC from a to b and wildcard-denial evidence. The attacker widens the interval to a through z, retaining other denial proofs. A resolver callout shows this evidence cached without validation, with its TTL subject to cache policy. A horizontal NTA-expiration bar states that validation resumes and unsafe continuation depends on unvalidated data surviving and being reused. In Phase 2, the client requests existing update with type A. The resolver finds update covered by a through z with supporting wildcard proof, fails to enforce validation state, and synthesizes false NXDOMAIN. A cross on the authority lifeline indicates no upstream query. The attacker may leave. The final callout limits denial to covered existing names and the remaining usable cache lifetime.}
\end{figure}

\subsection{Model Validation via Known Attacks}
\label{subsec:rediscovery}

We validated our model by confirming it identifies counter-examples for three known DNSSEC vulnerabilities.

\vspace{2pt} \noindent \textbf{Zone Enumeration.}
Our study reports violation of domain-name secrecy for NSEC (Property~16, Table~\ref{tab:verification-results}), consistent with its plaintext next-owner disclosure and known zone walking~\cite{wala2024zone}. The current artifact describes this exposure but lacks the executable P16 secrecy query, so it does not reproduce that reported check. P19 holds for NSEC3 only under the symbolic-name abstraction. It does not rule out dictionary attacks or the structured IPv6 reverse-zone enumeration discussed in Section~\ref{sec:discussion}.

\vspace{2pt} \noindent \textbf{Algorithm Downgrade Attack.}
\label{subsubsec:downgrade_attack}
Unsupported-algorithm handling can cause a resolver to skip validation and treat the delegation as \emph{Insecure}, allowing forged data without AD~\cite{heftrig2023downgrading}.
Property~13 (Table~\ref{tab:verification-results}) checks algorithm integrity. The model finds a counterexample where an on-path attacker changes the algorithm field to an unsupported value, causing acceptance of unvalidated records. Appendix~\ref{subsec:algo_scan} retains historical algorithm-support and signing measurements. Algorithm acceptance alone does not establish that this attack succeeds.

\vspace{2pt} \noindent \textbf{Cache Poisoning via CD Bit.}
\label{subsubsec:ruc_attack}
The RUCSEC attack~\cite{zhang2025your} exploits the Checking Disabled (CD) bit to bypass DNSSEC validation. Our model reveals that resolvers with a unified cache for validated and unvalidated responses are vulnerable. Property~7 (Table~\ref{tab:verification-results}) formalizes cache validation segregation and finds a valid attack path: an attacker sends a CD=1 query to inject an unsigned DNSKEY, and subsequent normal queries (CD=0) retrieve this unvalidated record, causing SERVFAIL responses.
\section{Root Causes and Recommendations}
\label{sec:rootcause}

\subsection{Specification-Level Analysis}
\label{subsec:spec_analysis}

The findings concern gaps in how RFC requirements interact. For P21, the wildcard-reconstruction procedure omits an explicit signer-depth lower bound, and the tested implementations share this omission. This does not imply that every conformant implementation must be vulnerable: additional containment checks can prevent the attack. Availability outcomes depend on the implementation.

\vspace{2pt} \noindent\ding{182} \textbf{Missing Zone-Cut Check on Wildcard Reconstruction.}
RFC 4034 \S3.1.3 requires only that the RRSIG \texttt{Labels} field not exceed the owner-name label count. RFC~4035 \S5.3.2 reconstructs the wildcard owner as ``\texttt{*.}'' prepended to the rightmost \texttt{Labels} labels and verifies the signature over that canonical form, without requiring the result to lie within the signer's zone. Because the attacker controls the child zone's key, the signature over the parent-zone wildcard form is cryptographically valid. Signature verification over that canonical form does not itself establish that the reconstructed owner is within the signer's authority. An explicit containment check is needed.

\vspace{2pt} \noindent\ding{183} \textbf{NTA Cache Hygiene Under-Specification.} RFC~7646's bypass lifecycle and RFC~8198's validation requirement leave post-expiry handling of NTA-cached records insufficiently explicit (Section~\ref{subsec:additional_findings}).

\vspace{2pt} \noindent\ding{184} \textbf{NSEC/NSEC3 Coexistence Under-Specification.} RFC~5155 permits transitions with both denial types. Safe reuse requires consistent proofs, while the tested fallback paths can combine inconsistent cached evidence (Section~\ref{subsec:additional_findings}).

\vspace{2pt} \noindent \textbf{Proposed RFC Updates.}

\vspace{2pt} \noindent \ding{182}~\textbf{RFC~4035}: Validators MUST reject an RRSIG where \texttt{Labels} $<$ \texttt{label\_count(Signer's Name)}. Equivalently, the reconstructed wildcard owner MUST be a subdomain of the Signer's Name.

\vspace{2pt} \noindent \ding{183}~\textbf{RFC~7646}: Resolvers SHOULD flush NSEC/NSEC3 records for the NTA subtree on NTA install and removal (including timer expiry). With this amendment added to our model, Property~23 verifies, i.e., NTA composition with aggressive caching is fixable.

\vspace{2pt} \noindent \ding{184}~\textbf{RFC~8198 / RFC~5155}: Resolvers SHOULD reject inconsistent denial proofs and MUST preserve proof validity during synthesis. This proposal allows legitimate NSEC/NSEC3 transitions while preventing reuse of contradictory evidence.

\vspace{2pt} \noindent \textbf{The Three Gaps in Context.}
RFC~6840 \S4.1 explicitly limits the zone scope of ancestor-side \emph{denial} records. The missing bound on reconstructed \emph{wildcard} owners motivates a corresponding containment check. Section~\ref{subsubsec:composition_survey} reviews these patterns.

Internet-Drafts for the three proposed updates and Errata for RFC~4034 \S3.1.3 and RFC~4035 \S5.3.2 remain planned.

\subsection{Implementation-Level Analysis}
\label{subsec:impl_analysis}

For NTA cache reuse, Unbound and Knot require each entry to have the RFC~4035 \emph{Secure} state before synthesis. PowerDNS and Technitium tests show weaker separation of unvalidated data. Hickory's non-validating-mode result does not exercise an NTA lifecycle. Validation-state checks alone do not supply the missing wildcard signer-depth bound in the tested Unbound and Knot versions. Section~\ref{subsec:additional_findings} and Table~\ref{tab:aggressive_vuln} distinguish these paths. Appendices~\ref{sec:code_analysis} and~\ref{app:additional_vulns} record the historical call sites.

\vspace{2pt} \noindent \textbf{Implementation Recommendations.}

\vspace{2pt} \noindent \ding{182}~\textbf{Bound \texttt{Labels} Below by Signer Depth} (Cross-Zone Wildcard Poisoning): At signature-verification time, reject any RRSIG whose \texttt{Labels} count is smaller than the label count of the Signer's Name as Bogus. This guard closes the modeled containment gap, but a complete implementation fix may also require cache-ordering and bailiwick checks, as Unbound's released patch demonstrates~\cite{unbound2026labels}. Operators should deploy the shipped vendor fixes (Section~\ref{subsec:pitfalls}).

\vspace{2pt} \noindent \ding{183}~\textbf{Track and Enforce Validation State} (NTA Cache Poisoning): revalidate or discard unvalidated entries before post-NTA reuse. \ding{184}~\textbf{Reject Inconsistent Denial Proofs} (Coexistence Attack): check the full proof and its zone context before cached synthesis.
\section{Related Work}
\label{sec:related}

\noindent\textbf{Formal DNS and DNSSEC modeling.}
Prior work has developed formal DNS models at both the standards and implementation levels. For authoritative nameservers, symbolic-execution approaches verify zone-file correctness~\cite{kakarla2020groot,kakarla2022scale}, implementation compliance via LLVM IR~\cite{zheng2023automated}, and CDN-scale deployments~\cite{larisch2024topaz}. For resolvers, DNS-Maude~\cite{liu2023formal} verifies caching and recursive query logic, and Nevatia~\cite{nevatia2024reachability} models communication processes with timers. On the DNSSEC side, early finite-state and theorem-proving efforts use Mur$\varphi$~\cite{bau2010security}, Isabelle/HOL~\cite{kammuller2014verification}, and Coq~\cite{eixarch2012formal}, but none model the cryptographic component that is central to DNSSEC. Chetioui et al.~\cite{chetioui2019formal} use ProVerif to verify that a resolver accepts only validly signed records, but model signing and verification of individual records in isolation, omitting resolver caching, zone delegation chains, and NSEC/NSEC3 denial-of-existence semantics. This single property cannot capture the cross-RFC interactions (e.g., RFC~7646 NTA with RFC~8198 aggressive caching) that give rise to the vulnerabilities we discover. The closest prior work, DNS-Maude~\cite{liu2023formal}, verifies DNS resolution comprehensively but excludes DNSSEC (``We do not reason about DNSSEC related behaviors'').

\vspace{2pt}\noindent\textbf{Positioning of our major finding.}
Ghost Domain Reloaded~\cite{li2023ghost} examines caching and delegation behavior that keeps revoked domains resolvable. MaginotDNS~\cite{li2023maginot} exploits bailiwick inconsistencies across forwarding and recursive modes sharing a cache. TuDoor~\cite{li2024tudoor} identifies logic flaws in response pre-processing with malformed packets.
Prior DNSSEC work~\cite{heftrig2023downgrading,heftrig2024harder,gruza2024attacking,daniluk2024zeros,zhang2025your} describes resource exhaustion~\cite{heftrig2024harder,gruza2024attacking,zhang2025your}, off-path poisoning~\cite{kaminsky2008black} after downgrade to unsigned resolution (AD$=$0)~\cite{heftrig2023downgrading}, and zone enumeration~\cite{wala2024zone,goldberg2015nsec5,borgolte2018enumerating}. Our comparison focuses specifically on attacker-chosen positive answers carrying AD$=$1, rather than all forms of forged DNS responses. The closest prior work is a pair of implementation bugs in NSEC processing, CVE-2017-15105~\cite{dolmans2018wildcard} and CVE-2018-1000003~\cite{powerdns2018ancestor}, in which specific resolvers accepted a signed NSEC record outside its intended scope and synthesized incorrect NXDOMAIN/NODATA. Cross-Zone Wildcard Cache Poisoning differs in three ways. First, it is a \emph{specification} gap (Section~\ref{subsec:spec_analysis}): the wildcard reconstruction rule lacks an explicit signer-depth bound shared by our tested implementations, whereas the earlier CVEs concerned specific denial-processing bugs. Second, it yields attacker-chosen \emph{positive} A records with AD$=$1 rather than denial of existing names. Third, it crosses the zone cut upward to poison some \emph{sibling} domains over which the attacker has no delegated authority. This attack uses the missing wildcard bound to obtain authenticated positive answers outside the signer's zone, without compromising an honest zone's key.
\section{Discussion}
\label{sec:discussion}

\noindent \textbf{Limitations.}
1) DNSSECVerif verifies abstract protocol models, not resolver code. Abstraction improves tractability, as in prior work~\cite{kakarla2022scale,liu2023formal}, but model-level guarantees do not establish implementation correctness. ResolverFuzz~\cite{zhang2024resolverfuzz} offers complementary implementation testing through query-response fuzzing and differential testing.
Trust anchors are static. KSK/ZSK rollover (RFC~6781), automated trust-anchor rollover (RFC~5011), algorithm rollover, and DS publication lag require additional state transitions. Extension groups E1--E4 concern trust-anchor lifecycles, parent-side CDS/DS updates, compact denial, and multi-epoch key sets (Appendix~\ref{app:rfc_survey}).
2) We represent Time-to-Live (TTL) using logical states such as active and expired, rather than concrete numerical values. This is sufficient for the cache-state properties studied here, but does not capture timing behavior at finer granularity.
3) Our adversary model follows the symbolic Dolev-Yao framework, which does not capture side-channel attacks or algorithmic complexity attacks. 
4) Property~19 abstracts away structured-name guessing. RFC~3596 \S2.5 represents IPv6 addresses under \texttt{ip6.arpa} as reversed hexadecimal labels~\cite{rfc3596}. Borgolte et al.~\cite{borgolte2018enumerating} use this structure and NSEC3 record types to enumerate IPv6 reverse zones through online collection and offline candidate hashing. Their attack combines RFC~5155 denial records with an operational namespace defined by another RFC. Modeling this namespace and adaptive candidate generation remains unimplemented extension E5. Our symbolic secrecy result does not bound its enumeration cost.
5) 
The survey excludes 15 edges concerning signaling, telemetry, registry-side, or out-of-band mechanisms, as summarized in Section~\ref{subsec:rfc_survey}. Application-layer uses such as DANE~\cite{rfc6698} also lie outside the modeled resolution process.
6) NSEC5~\cite{goldberg2015nsec5} uses verifiable random functions to prevent zone enumeration. We omit it because its Internet-Draft expired~\cite{draft-vcelak-nsec5}.

\vspace{2pt} \noindent \textbf{Generalization.} Our analysis can be applied to the 14 model-ready cross-RFC edges identified in Section~\ref{subsubsec:composition_survey}. These edges can be expressed using the current SAPIC+ model without adding new state or actors, and each follows one of the three patterns illustrated by Properties~21--23: zone-cut containment, denial-mechanism exclusivity, or preservation of validation state.

\vspace{2pt} \noindent \textbf{Does DNSSEC Reduce Security?}
An on-path attacker can already deny service to unsigned zones by dropping responses. P23 adds persistence after the bypass window if injected records survive and are reused. The reported BIND mixed-input crash has a separate availability consequence. P21 adds an authenticated-data integrity failure when a vulnerable resolver performs wildcard synthesis. Operators should deploy the shipped fixes while retaining DNSSEC validation. Model-level mitigation does not establish the security of an entire patched implementation.

\vspace{2pt} \noindent \textbf{Future Work.}
Future directions include extending the $\mathsf{ValidLabel}$ predicate to arbitrary label arithmetic to discover further zone-cut violations, code-level verification of defensive patterns via CBMC~\cite{kroening2014cbmc} or Frama-C~\cite{kirchner2015frama}, and longitudinal measurement of secure-pattern adoption across the resolver ecosystem.
\section{Conclusion}
\label{sec:conclusion}

In this paper, we present a cross-RFC formal analysis of DNSSEC, designed to analyze cross-RFC interactions that prior work examined only in isolation.
Using SAPIC+ models with Tamarin and ProVerif, we analyzed selected mechanisms from 10 DNSSEC-related RFCs and checked 23 properties. Seventeen hold under the modeled assumptions. Six are violated, exposing three new findings and three known attack classes.
Our models identify cross-zone wildcard cache poisoning and two further availability flaws (Section~\ref{subsec:additional_findings}). The missing wildcard signer-depth bound appears in four tested resolver implementations, with different serving outcomes. Tests of six resolvers distinguish malformed-record acceptance, authenticated forgery, and conditional denial of service.
Rediscovery of three known attack classes supports the model's treatment of those mechanisms.
Through this work, we provide a path toward strengthening both DNSSEC specifications and implementations.

\begin{acks}

We thank the reviewers for their insightful and constructive comments. 
We also appreciate Torsten Jahnke and Michael Silva Carvalho from Aviontex GmbH for sharing their experiences with aggressive caching and Negative Trust Anchors. 
This research is partially supported by NSF CNS-2047476, NSF CNS-2112471, the University of Texas System Rising STARs Award (No. 40071109), and the startup funding from the University of Texas at Dallas.
Any opinions, findings, and conclusions expressed in this material are those of the author(s) and do not necessarily reflect the views of the sponsors. 
OpenAI Codex and Anthropic Claude Code assisted with text editing and polishing.
\end{acks}

\appendix %
\section{Open Science} %
\label{app:open_science}

Our official artifact is prepared for release on GitHub\footnote{\url{https://github.com/DNSSECVerif/DNSSECVerif}}. It includes SAPIC+ models, ProVerif translations, verification scripts, and TLD census data. The authors' model and analysis code is licensed under the GPL-3.0 license.

The review artifact was provided through GitFront. The model import from source commit \texttt{c76c505} preserves 3,928 SAPIC+ source lines and four ProVerif companions. The repository records source provenance, model hashes, and a property-to-lemma map.

The artifact repository includes:
\begin{itemize}
    \item \itemtext{12 SAPIC+ model files (\texttt{.spthy}), including vulnerable and mitigation variants, associated with the properties in Table~\ref{tab:verification-results}: bounded cache checks (Properties~1--6, 8--9), authentication (Properties~10--12, 14), NSEC/NSEC3 denial, and cross-RFC interactions (Properties~21--23). The coverage gap for Property~16 is stated below.}
    \item \itemtext{4 ProVerif companion translations (\texttt{.pv}) for the NTA, algorithm-downgrade, and CD-bit models, alongside their SAPIC+ sources}
    \item Known attack validation models for algorithm downgrade (Section~\ref{subsubsec:downgrade_attack}) and CD-bit cache poisoning (Section~\ref{subsubsec:ruc_attack})
    \item \itemtext{The inherited verification script (\texttt{run\_all.sh}) and Docker build recipes, with execution limitations and per-property interpretation documented in the repository}
    \item \itemtext{The September~9 TLD census scripts, final classifications and exact lists, archived raw responses, integrity manifests, and a snapshot of the paper's TLD table}
\end{itemize}

Properties~8--9 are reachability checks. The retained NSEC model lacks an executable P16 secrecy query. Historical verification logs and resolver/PSL scan data are unavailable in this snapshot. The implementation exploit harness remains withheld for coordinated disclosure. Artifact preparation did not rerun verification or experiments, so the snapshot does not independently reproduce every reported result or timing.

The single-vantage census ran from 23:16 to 23:24 UTC on September~9, using DO-set queries through 1.1.1.1 and 8.8.8.8, with fallback checks for unresolved cases. The saved root-zone SOA serial is \texttt{2026090901}. The artifact retains 3,021 query observations, raw responses, timestamps, per-TLD classifications, input snapshots, and SHA-256 manifests.

\vspace{2pt} \noindent \textbf{TLD Record.}
\label{par:tld_record}
Table~\ref{tab:tld_nsec_census} lists the 52 authenticated NSEC-serving TLDs in our September~9, 2026 census. Except for \texttt{.gov}, each supplied a covering NSEC for our probe. These are TLD names, without establishing 52 distinct operator organizations or demonstrated compromises. Appendix~\ref{subsec:rfc8198_prevalence} gives the method and limitations.

\begin{table}[t]
\centering
\caption{All 52 TLDs serving NSEC with a root DS and AD-set responses in the September~9, 2026 census. Names use ASCII A-labels. The list records parent-zone observations, not demonstrated compromises.}
\label{tab:tld_nsec_census}
\small
\begin{tabular*}{\columnwidth}{@{\extracolsep{\fill}}lll@{}}
\toprule
\multicolumn{3}{c}{\textbf{TLDs with authenticated NSEC responses}} \\
\midrule
\texttt{.al} & \texttt{.aquarelle} & \texttt{.arpa} \\
\texttt{.audio} & \texttt{.auto} & \texttt{.ax} \\
\texttt{.bd} & \texttt{.br} & \texttt{.bt} \\
\texttt{.car} & \texttt{.cars} & \texttt{.ch} \\
\texttt{.christmas} & \texttt{.ci} & \texttt{.cy} \\
\texttt{.diet} & \texttt{.dz} & \texttt{.ee} \\
\texttt{.er} & \texttt{.flowers} & \texttt{.game} \\
\texttt{.gdn} & \texttt{.gn} & \texttt{.gov}$^\dagger$ \\
\texttt{.guitars} & \texttt{.hosting} & \texttt{.id} \\
\texttt{.kg} & \texttt{.kz} & \texttt{.lb} \\
\texttt{.li} & \texttt{.lk} & \texttt{.lol} \\
\texttt{.lr} & \texttt{.mc} & \texttt{.md} \\
\texttt{.mom} & \texttt{.nu} & \texttt{.pics} \\
\texttt{.pr} & \texttt{.ruhr} & \texttt{.se} \\
\texttt{.tc} & \texttt{.tn} & \texttt{.ve} \\
\texttt{.xn--54b7fta0cc} & \texttt{.xn--80ao21a} & \texttt{.xn--l1acc} \\
\texttt{.xn--mgbai9azgqp6j} & \texttt{.xn--mgbayh7gpa} & \texttt{.xn--pgbs0dh} \\
\texttt{.xn--xkc2al3hye2a} &  &  \\
\bottomrule
\end{tabular*}
\begin{minipage}{\columnwidth}
\footnotesize\raggedright
$^\dagger$\texttt{.gov} serves compact denial (NXNAME/TYPE128), not a covering NSEC range for our probe. The other 51 entries additionally have an authoritative covering-NSEC response, the parent-side prerequisite for P21. \texttt{.xn--fzc2c9e2c} also serves NSEC but has no root DS and is excluded from these 52. Exploitation still requires an attacker-controlled signed child, a suitable target name, and a vulnerable downstream resolver.
\end{minipage}
\end{table}

\section{Ethical Considerations} %
\label{app:ethics}

Stakeholders include resolver vendors, network operators, and DNS users. Vulnerability tests used only our isolated lab systems. Internet measurements used rate-limited, lightweight probes and a public contact for operator inquiries. We collected configuration data without personal information or exploit payloads. The September~9 census used ordinary DNS queries, limited to eight probe invocations per second, with additional authoritative checks. The study involved no human subjects.

We reported the findings to affected resolver vendors. Initial P21 reports to ISC, NLnet Labs, CZ.NIC, and PowerDNS were sent April~10--16, 2026. All four have shipped P21 fixes (Section~\ref{subsec:pitfalls}). Some vendors characterized the availability findings as design trade-offs. The PowerDNS response cited its documentation warning as mitigation. Standards updates remain planned.

On September~10, 2026, we emailed ICANN OCTO through its DNS-security reporting channel~\cite{icann-security-reporting}. We requested TLD-operator coordination to meet the Ethical Considerations requirements. ICANN created an internal tracking ticket. The dated TLD record appears in Appendix~\ref{app:open_science}. Operator notification and mitigation are not yet confirmed.

Publication can help operators assess exposure and support standards repairs. Deployed resolvers may remain unpatched, and some availability risks persist.
\section{Verification Results}
\label{sec:appendix_verification}

Table~\ref{tab:verification-results} in Section~\ref{sec:formal_models} reports all 23 properties. An \xmark\ in the Verified column indicates a violation. Auxiliary artifact checks cover reachability of cache initialization, authoritative responses, secure cache hits, and NTA bypass. Honest-signing-key secrecy is a separate safety check. These are not additional paper properties. The table retains study-reported outcomes and timings, not results regenerated during artifact preparation.

\section{Formalized DNSSEC model}
\label{sec:append_model}

The artifact contains SAPIC+ role and zone models and their included lemmas. Appendix~\ref{app:open_science} identifies the missing P16 query and retained-evidence limits. This appendix describes the encoding choices used in the main text.

\vspace{2pt} \noindent \textbf{Zone hierarchy.}
Zone files follow RFC~4035's signed-zone example: a root, an \texttt{example} TLD, and two SLDs, \texttt{a.example} and \texttt{b.example}. Separate models encode NSEC and NSEC3 behavior within this fixed hierarchy. The P21 witness instead uses selected names under \texttt{test.}, as described in Section~\ref{subsec:pitfalls}. Neither construction represents arbitrary zone contents.

\vspace{2pt} \noindent \textbf{Role processes.}
The client, authoritative-nameserver, and resolver roles carry the validation metadata described in Section~\ref{subsec:modeling_approach}. Response states distinguish \emph{Secure}, \emph{Insecure}, and \emph{Bogus} outcomes. Denial models distinguish NSEC, NSEC3, and mixed evidence. The coexistence model represents static, inconsistent views (Appendix~\ref{subsubsec:coexistence_attack}), while the NTA models separately represent bypass and expiry (Appendix~\ref{subsubsec:nta_vuln}). The bounded cache model, \texttt{cache\_bounded.spthy}, stores validated data as \emph{Secure} and returns \emph{Bogus} on validation failure. Its cache keys separate originating nameservers under entry-level \texttt{lock}/\texttt{unlock} semantics, making P3 structural. These models do not constitute one combined executable resolver.

\vspace{2pt} \noindent \textbf{Assumed cache interface.} The validation and denial models assume unexpired entries, preservation of validation state, and mutually exclusive cache access. Unvalidated data must pass validation before reuse as \emph{Secure}. Independent cache checks cover P1--6 and P8--9. P7 is violated in the CD-bit model, and P13 and P16 are also violated in the baseline set. These interface assumptions do not prove that all P10--20 hold after combining the model files. We provide no machine-checked refinement or composition theorem.

\begin{sloppypar}
\vspace{2pt} \noindent \textbf{RRSIG \texttt{Labels} handling.}
The released \texttt{rrsig\_underflow\_}\allowbreak\texttt{vulnerable.spthy} is a minimal witness model. Conditions 1--5 are recorded without independent guards, including the Labels upper bound and time validity. Condition~6 uses ground bailiwick facts, condition~7 checks a symbolic signature, and condition~8 is recorded after that check. Names and label counts are selected constants. The synthesis process requires the selected cached parent NSEC, but does not compute arbitrary interval coverage or closest-encloser proofs. Thus its counterexample motivates the packet-level checks in Section~\ref{subsec:pitfalls}, rather than proving conformance of a complete validator. The companion secure model excludes the modeled underflow case. This establishes the mitigation only within that abstraction.
\end{sloppypar}

Our model uses a fixed zone hierarchy. Its results depend on the encoded zone contents and denial proofs. In particular, P21 requires a suitable parent denial range. We do not exhaustively explore arbitrary zone files.

\section{Resolving Ambiguities}
\label{sec:append_ambiguities}

This appendix records the wildcard rule and cache-synchronization choices used in the model. Appendix~\ref{subsubsec:coexistence_attack} separately examines inconsistent denial proofs during NSEC/NSEC3 coexistence.

\subsection{Wildcard Handling with NSEC Records}
\label{subsec:ambiguity_wildcard}

RFC~4592 \S3.3.1 defines the wildcard source as \texttt{*.} followed by the closest encloser~\cite{rfc4592}. For \texttt{a.b.example.com}, if \texttt{b.example.com} is the closest encloser, the sole candidate is \texttt{*.b.example.com}. If that source is absent, the resolver does not fall back to \texttt{*.example.com}. The model applies this rule to both NSEC and NSEC3 alongside the signer-containment check in Section~\ref{subsec:auth_of_doe}. This is not a longest-match choice between multiple wildcard sources.

\subsection{Cache and Concurrency Ambiguities}
\label{subsec:ambiguity_cache_concurrency}

The DNS RFCs leave cache synchronization to implementers. RFC~1034~\cite[\S5.1]{rfc1034} introduces shared caches. RFC~1035~\cite[\S\S2.2, 6.1.2--6.1.3]{rfc1035} and RFC~2181~\cite[\S5.4.1]{rfc2181} specify TTL and ranking rules without synchronization primitives. RFC~4035~\cite[\S4.5]{rfc4035}'s atomic cache entry concerns record lifetimes, not mutual exclusion. Entry-level locking is therefore a modeling choice. P5 checks acquisition before release, and P6 checks store completion before a hit under the model restriction. The validation and NTA models separately track security context. Post-NTA reuse requires validation even if a cached entry remains unexpired (Appendix~\ref{subsubsec:nta_vuln}).

\section{Measurement Study about DNSSEC}
\label{sec:scanning}

This appendix retains historical deployment measurements alongside the dated TLD census. The September~2025 address inventory, May~2026 focused resolver scan, and September~2026 TLD census are separate cohorts. They characterize DNSSEC support and attack prerequisites, without measuring end-to-end exploitation or its prevalence. Historical resolver and signing-scan logs are unavailable in the retained artifact.

\subsection{Internet Scanning for DNSSEC-enabled Resolvers}
\label{subsec:resolver_scan}

We scanned the IPv4 address space using ZMap~\cite{durumeric2013zmap} (port 53, UDP/TCP) in September 2025, discovering 2,240,894 open resolvers. To identify DNSSEC-enabled resolvers, we queried each with a DNSSEC-signed test domain (Algorithm 13: ECDSA P-256/SHA-256 per RFC~6605~\cite[\S 4]{rfc6605}) and verified: (1)~\texttt{NOERROR} response, (2)~correct RRSIG in answer, and (3)~\texttt{AD} flag set.

\begin{table}[t]
\centering
\small
\caption{Geographic distribution of open resolvers and ASes (top 10 countries are listed).}
\label{tab:resolver_distribution}
\sisetup{
group-separator={,}, %
group-minimum-digits=4 %
}
\begin{tabular}{
l
S[table-format=7.0]  %
S[table-format=2.2]
S[table-format=5.0]
S[table-format=2.2]
}
\toprule
\textbf{Country} & {\textbf{Resolvers}} & {\textbf{(\%)}} & {\textbf{ASes}} & {\textbf{(\%)}}\\
\midrule
USA         & 647386 & 28.89 & 6331 & 16.56 \\
Germany     & 161644 &  7.21 & 1414 &  3.70 \\
France      & 145263 &  6.48 &  896 &  2.34 \\
Russia      & 111108 &  4.96 & 3363 &  8.79 \\
India       &  90501 &  4.04 & 1191 &  3.11 \\
Netherlands &  75829 &  3.38 &  835 &  2.18 \\
UK          &  65847 &  2.94 & 1040 &  2.72 \\
Japan       &  64433 &  2.88 &  619 &  1.62 \\
China       &  62009 &  2.77 &  273 &  0.71 \\
Canada      &  57259 &  2.56 &  806 &  2.11 \\
\midrule
Others      & 759615 & 33.89 & 21473 & 56.16 \\
\midrule
\textbf{Total}      & \textbf{\num{2240894}} & 100.00 & \textbf{\num{38241}} & 100.00 \\
\bottomrule
\end{tabular}
\end{table}

Table~\ref{tab:resolver_distribution} gives the geographic distribution: the resolvers span 222 regions with a long tail (the US hosts the largest share at 28.89\%).

\begin{table*}[htbp]
\small
\centering
\caption{Historical query-response counts by query type and status. The column sums differ from the reported address inventory.}
\label{tab:resolver_status}
\sisetup{group-separator={,}, group-minimum-digits=4}

\begin{tabular}{l *{4}{S[table-format=7.0]}}
\toprule
\begin{tabular}[c]{@{}l@{}}\textbf{Status and} \\ \textbf{Related RCODE}\end{tabular} & 
{\begin{tabular}[c]{@{}c@{}}\textbf{A record} \\ \textbf{w/o DNSSEC}\end{tabular}} & 
{\begin{tabular}[c]{@{}c@{}}\textbf{A record} \\ \textbf{w/ DNSSEC}\end{tabular}} & 
{\begin{tabular}[c]{@{}c@{}}\textbf{NSEC record} \\ \textbf{w/ DNSSEC}\end{tabular}} & 
{\begin{tabular}[c]{@{}c@{}}\textbf{NSEC3 record} \\ \textbf{w/ DNSSEC}\end{tabular}} \\
\midrule

Timeout & 0 & 1647 & 23615 & 9281 \\
NOERROR (0) & 360124 & 359373 & 96648 & 101697 \\
FORMERR (1) & 8920 & 10244 & 9034 & 8788 \\
SERVFAIL (2) & 41495 & 38717 & 139737 & 41777 \\
NXDOMAIN (3) & 12583 & 11160 & 150707 & 235972 \\
NOTIMP (4) & 3154 & 1247 & 3161 & 3164 \\
REFUSED (5) & 1786740 & 1789084 & 1789450 & 1784046 \\
YXDOMAIN (6) & 8 & 0 & 8 & 8 \\
NXRRSET (8) & 621 & 621 & 623 & 620 \\
NOTAUTH (9) & 38791 & 38783 & 38856 & 38798 \\
NOTZONE (10) & 18 & 5 & 18 & 18 \\
DSOTYPENI (11) & 1 & 1 & 1 & 1 \\
Unspecified Errors & 209 & 1782 & 806 & 28494 \\
\midrule

\multicolumn{5}{l}{\textbf{Sum of each query column:} \num{2252664}} \\
\bottomrule
\end{tabular}
\par\smallskip
\begin{minipage}{\textwidth}
\footnotesize\raggedright
The study reports 2,240,894 addresses. Each query column sums to 2,252,664, a difference of 11,770. The raw logs are unavailable, so these retained counts cannot be interpreted as unique-resolver totals.
\end{minipage}
\end{table*}

\subsection{DNSSEC Support by Open Resolvers}
\label{subsec:resolver_dnssec_support}

Table~\ref{tab:resolver_status} retains response counts by query type and status. Each column sums to 2,252,664, exceeding the stated address inventory of 2,240,894 by 11,770. Without the raw logs, we cannot determine the cause or interpret the columns as counts of unique resolvers. These probes also do not distinguish validation failures, configuration differences, or the attacks studied here. They therefore do not measure vulnerability prevalence.

\subsection{RFC~8198 Prevalence and NSEC-Signed Parent Zones}
\label{subsec:rfc8198_prevalence}

\noindent\textbf{Resolver-side measurement.} A separate focused scan on May~1, 2026 identified 143,186 open recursive resolvers by observing recursion to our authoritative server. The revised analysis counts 62,618 as validating: they accept the correctly signed test name and reject its mis-signed sibling. Aggressive NSEC synthesis was observed at 3,742 of these (6.0\%). The earlier, permissive definition included 3,769 additional resolvers that accepted mis-signed data. Those are excluded here. This cohort is separate from the September~2025 address inventory, and observed synthesis support is not an end-to-end exploit test or a measurement of factory defaults.

\vspace{2pt}\noindent\textbf{TLD-side measurement.} We repeated the parent-zone census on September~9, 2026, from 23:16 to 23:24 UTC. The seed is IANA's version \texttt{2026090900}, containing 1,438 TLDs~\cite{iana-tld-list}. The saved root-zone snapshot has SOA serial \texttt{2026090901}. For every TLD we queried fresh names with DO set through two independent recursive services (1.1.1.1 and 8.8.8.8), with a third service and direct authoritative checks for unresolved responses. We classify only denial records whose RRSIG signer is the measured TLD, so root-zone proofs and CNAME/DNAME targets are not misattributed. Wildcard answers can carry NSEC3 proofs without an SOA. Two apex queries for an unused private-use type resolved the remaining DNAME/wildcard cases. This was live DNS measurement, not a CZDS zone-file analysis.

All 1,438 TLDs were classified: 53 served NSEC, 1,309 served NSEC3, and 76 returned unsigned responses. These categories describe served records, not trust-chain status: 1,350 TLDs have a DS in the root snapshot. Among the NSEC group, 52 have a root DS and authenticated responses from both recursive services. Table~\ref{tab:tld_nsec_census} in Appendix~\ref{app:open_science} lists all 52. The remaining NSEC TLD, \texttt{.xn--fzc2c9e2c}, lacks a root DS and is not included in the authenticated set.

For 51 of the 52 authenticated NSEC TLDs, a direct query to a listed authoritative server also returned an authoritative NXDOMAIN with a TLD-signed NSEC covering the probe name. The exception is \texttt{.gov}: it returned compact denial with NOERROR, an NSEC owner equal to the queried name, and the NXNAME bit (TYPE128), consistent with RFC~9824~\cite{rfc9824}. This is not the covering-range proof used by Q2. Thus 52/1,438 (3.62\%) describes authenticated NSEC deployment, while 51/1,438 (3.55\%) is the set with the observed parent-side prerequisite for our trace.

These are dated observations from one vantage point, not proof of exploitation for every name or client. We did not test malicious payloads, assess each registry's registration policy, or measure the joint distribution of parent zones and vulnerable downstream resolvers. In particular, infrastructure zones such as \texttt{.arpa} do not imply unrestricted child registration. The artifact retains 3,021 query observations, raw responses, timestamps, input snapshots, all per-TLD classifications, and SHA-256 manifests. The prior July~23 public-suffix measurement reported 445 NSEC-serving apexes (315 ICANN and 130 private) among 7,097 PSL apexes. Its raw data are unavailable in this snapshot. It was not repeated in this TLD census and is not used to infer current exploitation prevalence.

\subsection{Measurement of Signing and Validation Algorithms}
\label{subsec:algo_scan}

These historical measurements describe algorithm acceptance and signing choices, not successful downgrade attacks (Section~\ref{subsubsec:downgrade_attack}). The validation table reports a cohort of 465,755 resolvers. Its relationship to the address and query-status totals cannot be reconciled without the raw logs. The signing study reports 3,904,220 Tranco-ranked domains~\cite{LePochat2019}, with 238,952 DS records from 203,395 domains and 530,055 DNSKEY records from 268,061 domains. The retained table's rows instead sum to 238,948 and 270,067. We preserve the historical counts and percentages while marking the unresolved units and denominators.

We annotate these historical measurements with the IANA DNS Security Algorithm Numbers registry as checked on September~10, 2026 (last updated August~10, 2026)~\cite{iana-dnssec-algorithms}. RFC~9904 supersedes RFC~8624 and makes this registry the canonical source of separate implementation and usage guidance for signing and validation~\cite{rfc9904}. Both tables retain the 14 measured algorithm IDs. Unmeasured registry entries are not assigned zero counts.

\begin{table*}[htbp]
\centering
\caption{DNSSEC Validation Success Rates by Algorithm}
\label{tab:dnssec_validation}
\sisetup{group-separator={,}}
\small %
\begin{tabular}{
    l
    S[table-format=2.0]
    l
    S[table-format=6.0]
    S[table-format=2.2]
    S[table-format=6.0]
    S[table-format=2.2]
    S[table-format=5.0]
    S[table-format=2.2]
}
\toprule
\multirow{2}{*}{\textbf{Algorithm Mnemonic}} &
\multicolumn{1}{c}{\multirow{2}{*}{\textbf{\#}}} &
\multicolumn{1}{c}{\textbf{IANA Status}} &
\multicolumn{2}{c}{\textbf{Responded (NOERROR)}} &
\multicolumn{2}{c}{\textbf{Resolved}} &
\multicolumn{2}{c}{\textbf{Validated}} \\
\cmidrule(lr){4-5} \cmidrule(lr){6-7} \cmidrule(lr){8-9}
& & \multicolumn{1}{c}{\scriptsize(Impl./Use for Validation)} & {\textbf{Count}} & {\textbf{(\%)}} & {\textbf{Count}} & {\textbf{(\%)}} & {\textbf{Count}} & {\textbf{(\%)}} \\
\midrule
\multicolumn{9}{l}{\textit{\textbf{Registry Use for Validation: RECOMMENDED}}} \\
RSASHA512       & 10 & MUST / REC & 327541 & 70.32 & 237487 & 50.99 & 46211 & 9.92 \\
ECDSAP256SHA256 & 13 & MUST / REC & 326998 & 70.21 & 236877 & 50.86 & 45342 & 9.74 \\
RSASHA256       & 8  & MUST / REC & 319828 & 68.67 & 229324 & 49.24 & 40499 & 8.70 \\
RSASHA1\textsuperscript{a} & 5 & MUST / REC & 336647 & 72.28 & 246915 & 53.01 & 17376 & 3.73 \\
ECDSAP384SHA384 & 14 & REC / REC  & 288712 & 61.99 & 197306 & 42.36 & 22362 & 4.80 \\
ED25519         & 15 & REC / REC  & 280908 & 60.31 & 189276 & 40.64 & 17174 & 3.69 \\
ED448           & 16 & REC / REC  & 319631 & 68.63 & 229234 & 49.22 & 13313 & 2.86 \\
RSASHA1-NSEC3-SHA1\textsuperscript{a} & 7 & MUST / REC & 238495 & 51.21 & 146206 & 31.39 & 78 & 0.02 \\
\midrule
\multicolumn{9}{l}{\textit{\textbf{Registry Use for Validation: MAY}}\textsuperscript{b}} \\
ECC-GOST12      & 23 & MAY / MAY  & 302912 & 65.04 & 208814 & 44.83 & 84    & 0.02 \\
SM2SM3          & 17 & MAY / MAY  & 302407 & 64.93 & 208017 & 44.66 & 87    & 0.02 \\
\midrule
\multicolumn{9}{l}{\textit{\textbf{Registry Use for Validation: MUST NOT}}} \\
DSA             & 3  & MUST NOT / MUST NOT & 337356 & 72.43 & 252575 & 54.23 & 1939 & 0.42 \\
RSAMD5          & 1  & MUST NOT / MUST NOT & 337409 & 72.44 & 247074 & 53.05 & 86 & 0.02 \\
DSA-NSEC3-SHA1  & 6  & MUST NOT / MUST NOT & 293356 & 62.99 & 202526 & 43.48 & 95 & 0.02 \\
ECC-GOST        & 12 & MUST NOT / MUST NOT & 298526 & 64.10 & 207680 & 44.59 & 80 & 0.02 \\
\bottomrule
\end{tabular}
\par\smallskip
\begin{minipage}{\textwidth}
\footnotesize\raggedright
\textbf{Note:} The table reports 465,755 resolvers and percentages relative to that cohort. Its relationship to the address and query-status totals remains unresolved without the raw logs. Statuses are IANA's implementation/use values for \emph{validation}, checked September~10, 2026 (registry updated August~10, 2026)~\cite{iana-dnssec-algorithms}. REC abbreviates RECOMMENDED. These are current annotations of historical measurements, not a new scan.
\par\textsuperscript{a}For algorithms~5 and~7, IANA retains MUST/RECOMMENDED for validation, but RFC~9905 \S2 also requires operators to treat them as unsupported, yielding Insecure responses without another supported validation path~\cite{rfc9905}. The registry labels alone are not deployment advice. \textsuperscript{b}SM2SM3 (RFC~9563~\cite{rfc9563}) and ECC-GOST12 (RFC~9558~\cite{rfc9558}) are defined by Informational, Independent Submission RFCs, not IETF Standards Track RFCs.
\end{minipage}
\end{table*}

\begin{table*}[htbp]
\centering
\caption{DNSSEC Algorithm Deployment in DS and DNSKEY Records}
\label{tab:dnssec_signing}
\sisetup{group-separator={,}}
\small
\begin{tabular}{
    l
    S[table-format=2.0]
    l
    S[table-format=6.0]
    S[table-format=2.2]
    S[table-format=6.0]
    S[table-format=2.2]
}
\toprule
\multirow{2}{*}{\textbf{Algorithm Mnemonic}} &
\multicolumn{1}{c}{\multirow{2}{*}{\textbf{\#}}} &
\multicolumn{1}{c}{\textbf{IANA Status}} &
\multicolumn{2}{c}{\textbf{DS (reported)}} &
\multicolumn{2}{c}{\textbf{DNSKEY (reported)}} \\
\cmidrule(lr){4-5} \cmidrule(lr){6-7}
& & \multicolumn{1}{c}{\scriptsize(Impl./Use for Signing)} & {\textbf{Count}} & {\textbf{(\%)}} & {\textbf{Count}} & {\textbf{(\%)}} \\
\midrule
\multicolumn{7}{l}{\textit{\textbf{Registry Use for Signing: RECOMMENDED}}} \\
ECDSAP256SHA256 & 13 & MUST / REC & 166736 & 69.78 & 200480 & 74.23 \\
RSASHA256       & 8  & MUST / REC & 57937  & 24.25 & 57297  & 21.22 \\
ED25519         & 15 & REC / REC  & 4289   & 1.79  & 2292   & 0.85  \\
\midrule
\multicolumn{7}{l}{\textit{\textbf{Registry Use for Signing: MAY}}} \\
ECDSAP384SHA384 & 14 & MAY / MAY  & 1753   & 0.73  & 1673   & 0.62  \\
ED448           & 16 & MAY / MAY  & 48     & 0.02  & 33     & 0.01  \\
SM2SM3          & 17 & MAY / MAY  & 0      & 0.00  & 0      & 0.00  \\
ECC-GOST12      & 23 & MAY / MAY  & 0      & 0.00  & 0      & 0.00  \\
\midrule
\multicolumn{7}{l}{\textit{\textbf{Registry Use for Signing: NOT RECOMMENDED or MUST NOT}}} \\
RSASHA1-NSEC3-SHA1 & 7 & NOT REC / MUST NOT & 3655 & 1.53 & 3762 & 1.39 \\
RSASHA512       & 10 & NOT REC / NOT REC  & 3389   & 1.42  & 3367   & 1.25  \\
RSASHA1         & 5  & NOT REC / MUST NOT & 1114   & 0.47  & 1154   & 0.43  \\
RSAMD5          & 1  & MUST NOT / MUST NOT & 11 & 0.00 & 0 & 0.00 \\
DSA             & 3  & MUST NOT / MUST NOT & 8 & 0.00 & 7 & 0.00 \\
ECC-GOST        & 12 & MUST NOT / MUST NOT & 6 & 0.00 & 1 & 0.00 \\
DSA-NSEC3-SHA1  & 6  & MUST NOT / MUST NOT & 2 & 0.00 & 1 & 0.00 \\
\bottomrule
\end{tabular}
\par\smallskip
\begin{minipage}{\textwidth}
\footnotesize\raggedright
\textbf{Note:} The study reports 3,904,220 domains, with 238,952 DS records from 203,395 domains and 530,055 DNSKEY records from 268,061 domains. The displayed rows instead sum to 238,948 and 270,067. Counts and percentages are retained unchanged. Count units (records or domains) and percentage denominators remain unresolved without the raw logs.
Statuses are IANA's implementation/use values for \emph{signing}, using the same September~10, 2026 snapshot as Table~\ref{tab:dnssec_validation}~\cite{iana-dnssec-algorithms}. REC and NOT REC abbreviate RECOMMENDED and NOT RECOMMENDED. Algorithm numbers identify the signature algorithm, including the algorithm field in DS records, not the DS digest type.
\end{minipage}
\end{table*}

\vspace{2pt} \noindent \textbf{Takeaways.}
Tables~\ref{tab:dnssec_validation} and~\ref{tab:dnssec_signing} show heterogeneous validation outcomes and signing concentrated on ECDSAP256SHA256 and RSASHA256. The retained measurements include RSASHA1-NSEC3-SHA1 and RSASHA1 (reported DS shares of 1.53\% and 0.47\%). RFC~9905 prohibits their use in new DNSKEY, RRSIG, and DS records~\cite{rfc9905}. Its operator requirements also qualify IANA's retained validation labels, so the 0.02\% validation rate for algorithm~7 alone does not establish noncompliance. DSA's measured 0.42\% validation rate and ECC-GOST's 0.02\% concern algorithms marked MUST NOT for validation in the dated registry snapshot. RFC~9906 retires ECC-GOST (12), not ECC-GOST12 (23)~\cite{rfc9906}. Historical observations are not present-day compliance tests, and aggregate choices do not establish whether a downgrade succeeds on an individual resolver.

\section{Source Code Analysis}
\label{sec:code_analysis}

This appendix provides per-resolver source-code analysis for the headline RRSIG \texttt{Labels} underflow (Section~\ref{subsec:pitfalls}). Per-resolver call-site detail for the two DoS-class findings appears with each finding in Appendix~\ref{app:additional_vulns}. Resolver-internal symbols quoted below (\texttt{sec\_} \texttt{status\_*}, \texttt{KR\_RANK\_*}, \texttt{vState::*}) map directly to the RFC~4035 \S4.3 \emph{Secure}/\emph{Insecure}/\emph{Bogus} vocabulary.

\subsection{RRSIG Labels Underflow}
\label{subsec:rrsig_code_analysis}

All four tested implementations with RFC~8198 wildcard synthesis omit the signer-depth lower bound. This shared omission motivates an explicit specification-level containment rule, but serving outcomes and the full implementation repairs differ. The following code locations describe the historical tested versions.

\vspace{2pt} \noindent \textbf{BIND 9.20.22.}
\texttt{dnssec.c:450} computes the canonical owner by splitting the name at \texttt{sig.labels+1} and returns the \texttt{FROMWILDCARD} result code without testing that the result is a subdomain of the signer name. The reconstructed owner is then handed to \texttt{rbtdb.c} which inserts it into the synth-from-DNSSEC tree at \emph{Secure} trust. The containment guard rejects reconstructed owners outside the signer's zone.

\vspace{2pt} \noindent \textbf{Unbound 1.24.2.}
\texttt{val\_sigcrypt.c:1650} reads \texttt{labels = sig[3]} and proceeds to canonicalization without comparing against the signer's label count. Separately, \texttt{validator.c:1036} adds the wildcard to the negative cache before \texttt{validate\_msg()} completes, so the poisoned owner is cached even if a later check failed. The released fix adds the lower bound, delays cache writes until validation completes, and checks wildcard bailiwick~\cite{unbound2026labels}.

\vspace{2pt} \noindent \textbf{Knot Resolver 6.2.0.}
\texttt{lib/cache/api.c:561} computes \texttt{wild\_labels = owner\_labels - rrsig\_labels} without bounding it against the owner-minus-signer distance, so \texttt{knot\_dname\_next\_labels()} walks past the zone cut and the resulting \texttt{kr\_cache\_key\_wildcard()} key lands at \texttt{*.test.}
The released fix rejects RRSIGs whose \texttt{Labels} count is smaller than the signer's depth~\cite{knot2026dnssec}.

\vspace{2pt} \noindent \textbf{PowerDNS Recursor 5.4.0.}
\texttt{dnssecinfra.cc:548} reconstructs \texttt{realOwner} from \texttt{Labels} and \texttt{syncres.cc:4884} writes it into the aggressive NSEC cache \emph{before} \texttt{checkWildcardProof()} runs. The correction at \texttt{syncres.cc:5001} repairs only \texttt{qname}, not \texttt{realOwner}. The repair requires signer containment for ANSWER-section RRSIGs and consistent validation state for \texttt{realOwner}.

\vspace{2pt} \noindent \textbf{Technitium and Hickory (Historical Phase~1 Only).}
Both tested versions accept a child-signed RRSIG whose reconstructed wildcard owner lies above the zone cut (Phase~1), but neither implements RFC~8198 wildcard synthesis (Phase~2). Technitium uses RFC~2308 negative caching only (\texttt{DnsCache.cs:32}). Hickory's response cache (\texttt{cache.rs:48}) stores whole responses keyed by \texttt{Query}, not reusable wildcard entries.

\section{Cross-RFC Survey: Supporting Listings}
\label{app:rfc_survey}

This appendix supplements Section~\ref{subsec:rfc_survey} with 14 model-ready and 13 extension entries, plus exclusion rationales. The table does not enumerate all 52 edges, and model-ready entries are not additional verified results. A per-model RFC-section index (mapping each SAPIC+ model to the RFC clauses it captures, with abstractions and omitted behavior) is provided in the artifact's \texttt{MODEL\_INDEX.md}.

\vspace{2pt} \noindent \textbf{Corpus retrieval.}
The survey used IETF Datatracker document lists and canonical RFC text from \texttt{rfc-editor.org}. Its selected 41-RFC corpus covers \texttt{dnsext}, \texttt{dnsop}, and related publication streams, rather than 41 RFCs all published by those two working groups. The paragraph filter uses \emph{validator}, \emph{resolver}, \emph{cache}, \emph{trust anchor}, \emph{Secure}, \emph{Insecure}, \emph{Bogus}, \emph{Indeterminate}, and DNSSEC RR-type names. The reviewer-added RFC~5155/RFC~3596 edge is separate from the original corpus.

\vspace{2pt} \noindent \textbf{Per-edge listing.}
Table~\ref{tab:rfc_interactions_full} lists every model-ready and extension edge with its pattern (Section~\ref{subsubsec:composition_survey}) or extension group (E1--E5).

\begin{table*}[t]
\centering
\caption{The survey's 14 model-ready and 13 extension edges, grouped by failure pattern or required extension. Covered edges are omitted, and Appendix~\ref{app:rfc_survey} gives exclusion rationales. These entries are analysis candidates, not additional verified results.}
\label{tab:rfc_interactions_full}
\footnotesize
\setlength{\tabcolsep}{3.5pt}
\begin{tabular*}{\textwidth}{@{\extracolsep{\fill}}llQ{9.2cm}l@{}}
\toprule
\textbf{RFC$_a$ $\times$ RFC$_b$} & \textbf{\S} & \textbf{Informal constraint} & \textbf{Pattern} \\
\midrule
\multicolumn{4}{@{}l}{\emph{Zone-cut containment} (anchor: Property~21)} \\
6840 $\times$ 4035 & 6840 \S4.1  & Ancestor-side NSEC/NSEC3 proves nothing about names below the zone cut. & P21-family \\
4470 $\times$ 8198 & 4470 \S3    & Minimally-covering NSEC ($\varepsilon$-range) permits no synthesis for a name $\neq$ owner. & P21-family \\
4470 $\times$ 4035 & 4470 \S3    & Minimal-cover NSEC type bitmap is exactly $\{$RRSIG, NSEC$\}$. & P21-family \\
6840 $\times$ 5155 & 6840 \S5.10 & Opt-out NSEC3 does not prove nonexistence of an insecure delegation. & P21-family \\
\addlinespace[2pt]
\multicolumn{4}{@{}l}{\emph{Denial-mechanism exclusivity} (anchor: Property~22)} \\
8198 $\times$ 9077 & 9077 \S3.4  & Synthesized-answer TTL $\leq \min(\text{NSEC TTL}, \text{SOA minimum})$. One TTL epoch per range. & P22-family \\
6840 $\times$ 4035 & 6840 \S4.3  & NODATA proof requires CNAME bit clear in the covering NSEC bitmap. & P22-family \\
6840 $\times$ 4035 & 6840 \S5.11 & Any single valid algorithm path suffices for \emph{Secure} (assumption of P10/P14). & P22-family \\
\addlinespace[2pt]
\multicolumn{4}{@{}l}{\emph{Validation-state preservation} (anchor: Property~23)} \\
9276 $\times$ 8198 & 9276 \S3.2  & Iteration-downgraded (Insecure) NSEC3 is not used for aggressive synthesis. & P23-family \\
5155 $\times$ 9276 & 9276 \S3.2  & Valid-signature NSEC3 with iteration count above threshold never yields state \emph{Secure}. & P23-family \\
8624 $\times$ 4035 & 8624 \S3.3  & Zone signed only with a MUST-NOT-validate algorithm $\Rightarrow$ \emph{Insecure} $\land$ AD$=$0. & P23-family \\
8020 $\times$ 4035 & 8020 \S2    & RFC~8020 NXDOMAIN-cut applied only to a DNSSEC-validated NXDOMAIN. & P23-family \\
8020 $\times$ 8198 & 8020 \S2    & NXDOMAIN-cut at $n$ agrees with any cached NSEC range covering $n$. & P23-family \\
8767 $\times$ 4035 & 8767 \S4    & Serve-stale data carries AD$=$0 unless its RRSIG is still temporally valid. & P23-family \\
8767 $\times$ 8198 & 8767 \S4    & Expired NSEC is not used for aggressive synthesis. & P23-family \\
\addlinespace[2pt]
\cmidrule{1-4}
\multicolumn{4}{@{}l}{\emph{Require model extension}} \\
5011 $\times$ 4035 & 5011 \S2.2  & No window during TA hold-down where neither old nor new TA validates the apex DNSKEY. & E1 \\
5011 $\times$ 4035 & 5011 \S4.2  & Revoked key is permanently removed from the TA set. & E1 \\
5011 $\times$ 6840 & 6840 \S5.3  & Setting REVOKE alters the key tag. DS$\leftrightarrow$DNSKEY matching must recompute. & E1 \\
7344 $\times$ 4035 & 7344 \S6.1  & Every parent-published DS matches some live child DNSKEY (no CDS-poll race). & E2 \\
8078 $\times$ 4035 & 8078 \S4    & DS deletion (CDS$=$0) only after Secure-validated signal. & E2 \\
9615 $\times$ 8078 & 9615 \S4.2  & DS install only if the signaling-name chain is Secure. & E2 \\
9615 $\times$ 4035 & 9615 \S5.2  & Stale cached CDS is not used for DS install. & E2 \\
9824 $\times$ 4035 & 9824 \S3.1  & NXNAME sentinel + NOERROR is semantically NXDOMAIN. & E3 \\
9824 $\times$ 8198 & 9824 \S3    & Compact-DoE NSEC is not range-synthesized. It is mapped to NXDOMAIN before caching. & E3 \\
9824 $\times$ 8020 & 9824 \S3.1  & NOERROR containing NXNAME triggers RFC~8020 cut iff resolver recognizes the sentinel. & E3 \\
6781 $\times$ 4035 & 6781 \S4.1  & KSK/ZSK pre-publish and double-signature rollovers never open a Bogus window. & E4 \\
8901 $\times$ 6840 & 8901 \S2    & Multi-signer zone is \emph{Secure} iff some provider path validates. & E4 \\
5155 $\times$ 3596 & 3596 \S2.5 & NSEC3 hashes over the structured IPv6 reverse namespace permit candidate enumeration. Extending P19 requires modeling that namespace and the adversary's guesses~\cite{borgolte2018enumerating}. & E5 \\
\bottomrule
\end{tabular*}
\end{table*}

\vspace{2pt} \noindent \textbf{Model extensions (E1--E5).}
\emph{E1, RFC~5011 trust-anchor state machine}: a per-TA state (Start / AddPend / Valid / Missing / Revoked / Removed) with hold-down timers. Candidate properties: no window where neither old nor new TA validates the apex DNSKEY, and a revoked key is permanently removed from the TA set. \emph{E2, CDS/DS lifecycle (RFC~7344/8078/9615)}: a parent-side polling agent that reads child CDS/CDNSKEY and mutates the parent DS RRset. Candidate properties: every published DS matches some live child DNSKEY, and DS deletion (CDS$=$0) requires a Secure-validated signal. \emph{E3, RFC~9824 compact denial}: an NXNAME pseudo-type and an RCODE-rewriting step on the resolver side. Candidate property: compact-DoE NSEC is never range-synthesized and is mapped to NXDOMAIN before RFC~8020 cut. \emph{E4, multi-epoch key sets (RFC~6781/8901)}: time-indexed DNSKEY/DS sets with overlapping validity. Candidate property: pre-publish, double-signature, and multi-signer rollovers never open a Bogus window. \emph{E5, operational zone structure (RFC~5155 $\times$ RFC~3596)}: extend P19 with the nibble-tree namespace of \texttt{ip6.arpa}, record-type information, and candidate generation and hashing, following Borgolte et al.~\cite{borgolte2018enumerating}. This reviewer-suggested edge lies beyond the original 41-RFC corpus. We have not implemented the extension or analyzed enumeration cost.

\vspace{2pt} \noindent \textbf{Out of scope.}
RFC~8145/8509 (trust-anchor telemetry, no validator-state change), RFC~8914 (extended error codes, post-hoc reporting only), RFC~6975 (algorithm-understanding signaling), RFC~7477/8976 (CSYNC and ZONEMD, acted on by registry or zone-transfer recipient, not the validating resolver), RFC~9157/9904 (IANA process, no validator behavior), RFC~7583 (informational timing parameters), RFC~8027 (operational fallback heuristics, SHOULD/MAY only), RFC~8749 (DLV removal, pure simplification), RFC~7958/9718 (out-of-band root TA distribution, bootstraps the model's initial state), and the signer-side TTL clauses of RFC~9077 \S3.1/3.3 (the resolver receives whatever is sent, the resolver-side bound is under P23).

\section{Additional Cross-RFC Vulnerabilities}
\label{app:additional_vulns}

This appendix adds attack and implementation details for the two availability findings in Section~\ref{subsec:additional_findings}. The modeled scenarios require an NTA or inconsistent mixed denial records. The separately reported BIND crash needs only attacker-served mixed input. These are availability findings, distinct from the P21 positive-answer forgery.

\vspace{2pt} \noindent \textbf{Attacker Models.}
P23 requires an on-path attacker during an operator-enabled NTA window. The modeled P22 case requires usable, inconsistent denial evidence and an affected resolver path. The illustrative CDN scenario below does not establish that an arbitrary co-tenant can supply those proofs. The separate BIND crash uses attacker-served mixed input. False denial affects clients whose requested names are covered by the reused proofs. Technitium's P23 result concerns ordinary negative caching, not RFC~8198 synthesis.

\subsection{NTA Cache Poisoning}
\label{subsubsec:nta_vuln}

\noindent \textbf{Root Cause.} %
An NTA temporarily bypasses DNSSEC validation for a configured subtree. When it expires, validation resumes, but previously cached records may remain. P23 arises if unvalidated denial records survive with usable lifetimes and are reused without validation. This conflicts with RFC~8198's requirement for validated synthesis evidence. Expiry alone does not establish that every cache removes or revalidates such records.

\vspace{2pt} \noindent \textbf{Attack Scenario.}
Figure~\ref{fig:nta_attack} illustrates a conditional attack: an on-path attacker injects unvalidated NSEC records during an NTA window. False denial after expiry requires those records to remain cached with usable lifetimes and to pass a synthesis path without a validation-state check. The formal models abstract lifetimes and do not establish that every configured NTA or cache has these retention properties.

\vspace{2pt} \noindent \textbf{Real-World Instance.}
During the May~2026 \texttt{.de} DNSSEC outage, Cloudflare applied a validation override functionally similar to an NTA~\cite{cloudflare2026detld}. This documents operational use of validation bypass at TLD scope. It does not establish that Cloudflare retained unvalidated denial records after restoring validation or was vulnerable to P23. Those additional conditions were studied in our models and controlled experiments.

\vspace{2pt} \noindent \textbf{Formal Verification.}
Property~23 (Section~\ref{subsubsec:further_composition}) requires validation state preservation. The vulnerable SAPIC+ model includes an ordered \texttt{attack\_with\_nta\_expiry} reachability lemma. Its ProVerif companion queries denial, injection, caching, and synthesis events. Event co-occurrence alone does not establish the ordered expiry trace. The artifact's property map distinguishes these checks.

\vspace{2pt} \noindent \textbf{Testing Resolver Software.}
Table~\ref{tab:aggressive_vuln} summarizes outcomes. The protective pattern is a per-entry validation-state check before synthesis: Unbound enforces \texttt{sec\_status\_secure} at four points in \texttt{val\_neg.c}. Knot requires \texttt{KR\_RANK\_SECURE} (rank~32) which NTA zones (rank~8) cannot reach. BIND is partial: explicit \texttt{rndc nta -remove} flushes the cache but timer expiry (\texttt{nta.c:delete\_expired()}) does not. The PowerDNS and Technitium tests show reuse of unvalidated cached data. Hickory's experiment disables validation globally. Because Hickory has no NTA lifecycle, it is not evidence of persistence across NTA expiry. Technitium's result concerns ordinary negative caching, since it does not implement RFC~8198 synthesis.

\subsection{NSEC/NSEC3 Coexistence Attack}
\label{subsubsec:coexistence_attack}

\begin{figure}[!tbp]
\centering
\includegraphics[page=5,width=\columnwidth]{figures.pdf}
\caption{NSEC/NSEC3 coexistence scenario during a CDN migration. Phase~1 caches both denial types. Phase~2 falls back to NSEC3 for a victim query. False denial requires inconsistent evidence or a proof-processing flaw, not hash coverage alone.
}
\label{fig:mix_nsec_nsec3}
\Description{Two-phase swimlane diagram with client, resolver, and CDN-authority lifelines and thirteen numbered steps. A heading identifies a schematic CDN migration between NSEC and NSEC3. During cache priming, an A query for aa.cdn.net obtains NXDOMAIN with NSEC from a.cdn.net to b.cdn.net. A second A query, for x.cdn.net outside that interval, obtains NXDOMAIN with NSEC3 denial proof records. Two cache callouts show both denial types retained and state that proof validity is still required. In Phase 2, an A query for c.cdn.net, identified as an existing victim name, follows three resolver steps: the NSEC interval misses. The resolver evaluates the full NSEC3 denial proof. It synthesizes NXDOMAIN without an upstream query. An explicit condition inside this processing callout states that false denial requires inconsistent evidence or a proof-processing flaw. The synthesized response returns to the client, and the final callout labels the denial conditional and limited to the usable lifetime of the cached proof. Concrete signed proof records and an anycast attack trace are not supplied by this schematic.}
\end{figure}

\noindent \textbf{Root Cause.}
An affected resolver may try NSEC3 after cached NSEC fails to cover a query. NSEC and NSEC3 use different orderings, but this difference alone does not imply incorrect denial. RFC~5155 permits NSEC/NSEC3 transitions during which both types coexist. False denial additionally requires usable inconsistent proofs or a proof-processing flaw. Proof validity, rather than the presence of two denial types, is the relevant condition.

\vspace{2pt} \noindent \textbf{Attack Scenario.}
Figure~\ref{fig:mix_nsec_nsec3} illustrates a potential CDN or managed-DNS transition scenario: queries populate a resolver's cache from NSEC and NSEC3 views. Incorrect denial additionally requires inconsistent, usable signed proofs and an affected synthesis path. Falling outside an NSEC interval and inside an NSEC3 hash interval alone is insufficient, since closest-encloser and other denial checks still apply. The static formal model does not establish a concrete CDN co-tenancy trace or the prevalence of such inconsistent views.

\vspace{2pt} \noindent \textbf{Formal Verification.}
Property~22 (Section~\ref{subsubsec:further_composition}) checks consistency of existence and denial outcomes for the same modeled name. The \texttt{denial\_exclusivity} lemma compares those events in static NSEC/NSEC3 views. It is not a one-denial-type-per-cache invariant or a full implementation of a signed NSEC3 denial proof.

\vspace{2pt} \noindent \textbf{Testing Resolver Software.}
Table~\ref{tab:aggressive_vuln} summarizes controlled mixed-response tests. Knot's two-slot cache (\texttt{cache/impl.h}) tries NSEC and NSEC3 sequentially and produced false denial. PowerDNS's result depends on asymmetric zone-type transitions. Hickory's default non-validating mode bypasses mixed-record rejection, so its result is not equivalent to a validating resolver's authenticated-denial failure. Unbound lacks general NSEC3 synthesis and does not exhibit this fallback. Separately, attacker-served mixed input triggers a \texttt{RUNTIME\_CHECK} abort in BIND~9.20.22 under its default configuration. That crash was disclosed to ISC and does not require an honest mixed deployment.
Technitium does not implement RFC~8198.

The query-status measurements in Appendix~\ref{subsec:resolver_dnssec_support} do not test coexistence attacks or identify naturally mixed zones. They therefore cannot quantify P22 prevalence or its exploitation rate.

\end{document}